\documentclass[9pt,twocolumn]{extarticle}
\usepackage{amsmath, amsthm, amssymb}
\usepackage{graphicx}
\usepackage{textcomp}
\usepackage{times}
\usepackage{textsub}
\usepackage{stmaryrd} 
\usepackage[normalem]{ulem}
\usepackage{fix-cm} 
\usepackage[none]{hyphenat}
\usepackage[]{caption}
\usepackage{pdfpages}
\DeclareCaptionLabelSeparator{justperiod}{.\ }
\usepackage{float}
\usepackage{wrapfig} 
\usepackage[version=3]{mhchem}
\usepackage{bm}
\usepackage{tikz} 
\usepackage{dblfloatfix} 
\usepackage{fixltx2e} 
\usepackage{xcolor}

\usepackage[numbers,round,sort&compress]{natbib} 

\def\fonesize{7in}

\newcommand{\bsf}[1]{{\bfseries\sffamily {#1}}}

\renewcommand{\vec}[1]{\mathbf{#1}}
\newcommand{\Mypm}{\mathbin{\tikz [x=1.4ex,y=1.4ex,line width=.2ex] \draw (0.0,0) -- (1.0,0) (0.5,0.08) -- (0.5,0.92) (0.0,0.5) -- (1.0,0.5);}}%

\newcommand{\fx}[1]{\textcolor{black}{#1}}

\def\svsp{\vspace*{0.05in}}

\def\um{$\mu$m}

\def\degree{$^\circ$}
\def\degreebf{$\bm{^\circ}$}

\def\figcaptionbaselinestretch{1.1}

\title{ {\bfseries\sffamily \fontsize{25.0}{29.16}\selectfont 
\vspace{-.50in} 
\hspace*{-1.56201825in} Absolute and arbitrary orientation \\
\hspace*{-2.80780910505in} of single molecule shapes \\
}}

\author{ \hspace*{-2.45744205258473in} {\bfseries\sffamily \begin{tabular}{l l l} 
   Ashwin Gopinath$^{\bm{1}}$ & Chris Thachuk$^{\bm{2}}$ & Anya Mitskovets$^{\bm{3}}$ \\
   Harry A. Atwater$^{\bm{3}}$ & David Kirkpatrick$^{\bm{5}}$ & Paul
    W. K. Rothemund$^{\bm{1,2,4,}}$\thanks{Departments of
      $^{1}$Bioengineering, $^{2}$Computing \& Mathematical Science,
      $^{3}$Applied Physics \& Materials Science, $^{4}$Computation \&
      Neural Systems, California Institute of Technology, Pasadena, CA
      91125, USA and $^{5}$Department of Computer Science, University
      of British Columbia, Vancouver, Canada.\newline $^*$ Address
      correspondence to: ashwing@caltech.edu, pwkr@dna.caltech.edu} \\
    \end{tabular}}
\vspace*{-0.375in}
 }
\date{}  
\begin{document}
\maketitle

 {\bsf{\noindent DNA origami is a modular platform for the combination
     of molecular and colloidal components to create optical,
     electronic, and biological devices. Integration of such nanoscale
     devices with microfabricated connectors and circuits is
     challenging: large numbers of freely diffusing devices must be
     fixed at desired locations with desired alignment. We present a
     DNA origami molecule whose energy landscape on lithographic binding sites
     has a unique maximum. This property enables device align- ment
     within $\Mypm$3.2\degreebf\ on SiO\textsubscript{2}. Orientation is
     absolute (all degrees of freedom are specified) and arbitrary
     (every molecule's orientation is independently specified).  The
     use of orientation to optimize device performance is shown by
     aligning fluorescent emission dipoles within microfabricated
     optical cavities. Large-scale integration is demonstrated via an
     array of 3,456 DNA origami with 12 distinct orientations, which
     indicates the polarization of excitation light.}}

\begin{figure*}[!hb]    
\centerline{ \includegraphics[width=\fonesize]{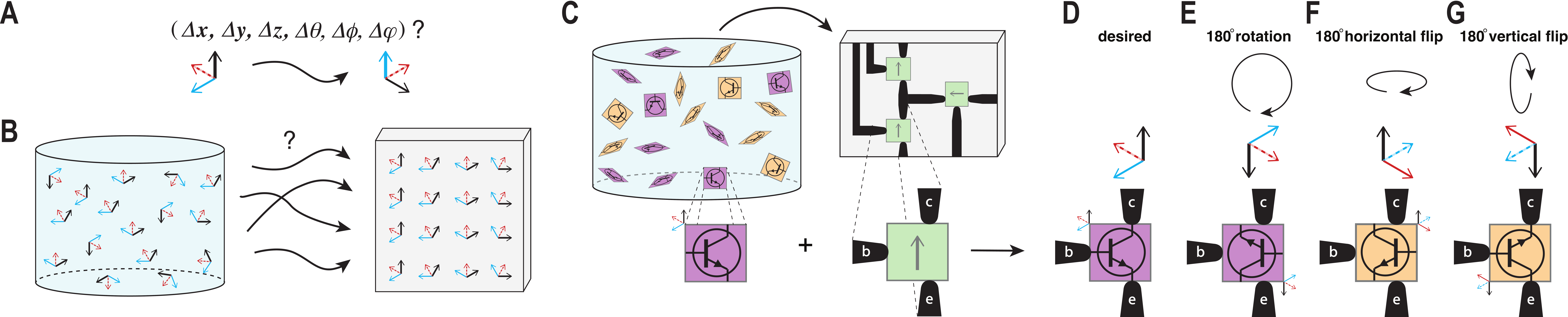} }  \vspace{-.1in}
\renewcommand{\baselinestretch}{\figcaptionbaselinestretch}
\caption{\bsf{Challenges for DSA of origami-templated devices.}
  (\textbf{A}) The mathematical problem of absolute orientation. Bold
  arrows show in-plane axes, dotted arrows point into page; regular
  arrows point out. (\textbf{B}) The physical problem of absolutely
  orienting solution-phase (blue) devices on planar substrates (gray)
  so that each device has an arbitrary, user-specifiable
  orientation. (\textbf{C}) DOP scheme for an asymmetric device
  (bipolar junction transistor) shows the problem of using high
  symmetry origami. Rectangles would attach to binding sites (green)
  with four orientations (\textbf{D}--\textbf{G}), two right-side up
  (purple) and two up-side down (orange). Electrodes c, e, and b can
  only connect to transistor collector, emitter, and base in a single
  (desired) orientation. Global methods are ruled out: coordinate
  systems attached to origami indicate symmetries that prevent fields
  or flow
  from distinguishing D--G; the intended circuit contains three sites
  and two orientations (gray arrows) requiring arbitrary orientation.}
\vspace{-.1in}
\label{fig:fig1}
\end{figure*} 

\begin{figure*}[!hb]    
\centerline{ \includegraphics[width=\fonesize]{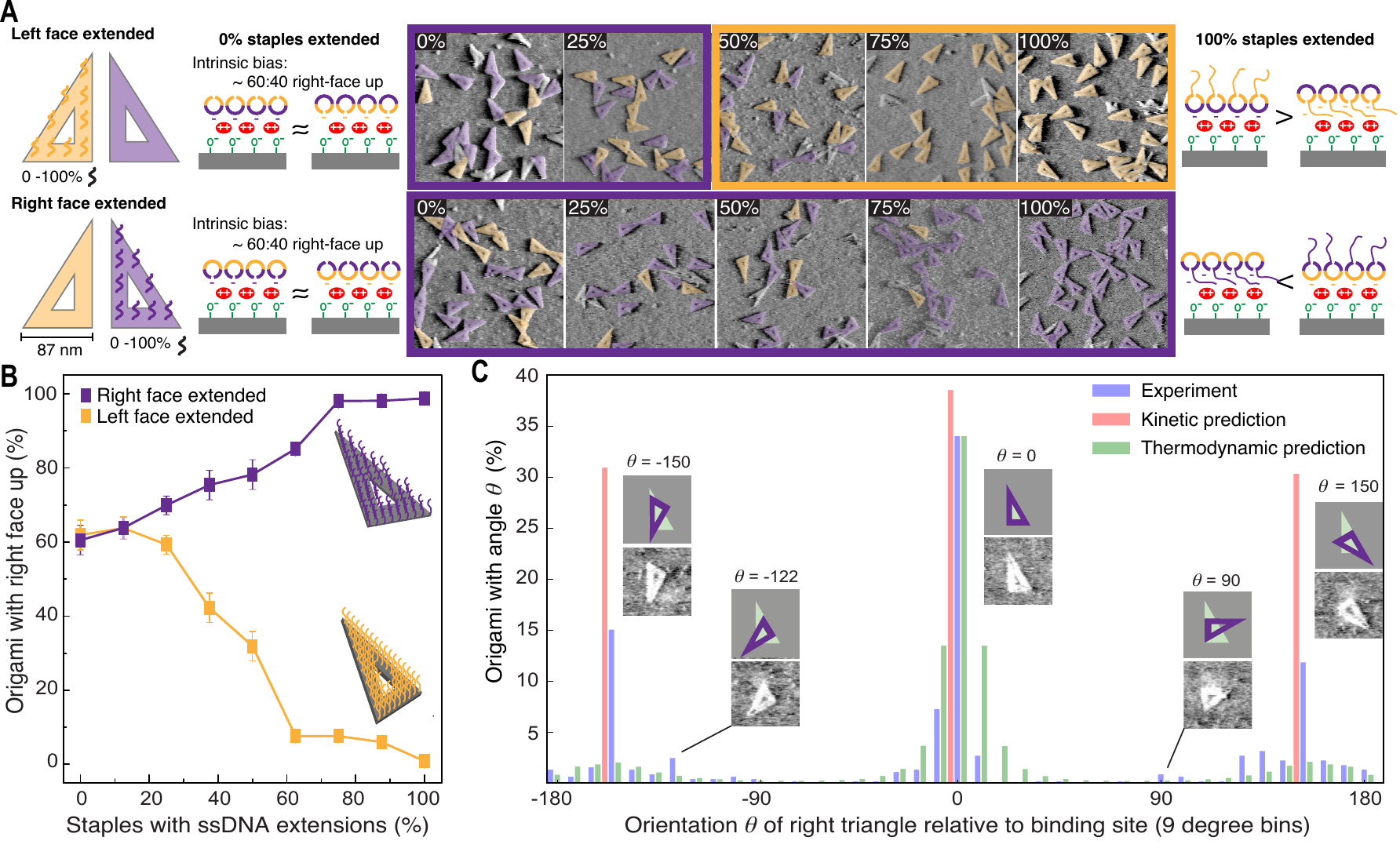} }  \vspace{-.1in}
\renewcommand{\baselinestretch}{\figcaptionbaselinestretch}
\caption{\bsf{DOP of asymmetric right triangles.} (\textbf{A}) Before
  DOP, up-down symmetry was broken by extending staples on either the
  right face (purple) or left face (orange) with ssDNA (20 nt,
  poly[T]) at nicks in the phosphate backbone. From zero (0\%) to 200
  staples (100\%) were extended. AFM shows that 100\% extension caused
  nearly 100\% bias on unpatterned SiO\textsubscript{2}, with
  extensions facing up; outline color reflects bias. Extensions
  interfere with binding of negatively-charged phosphate groups to
  Mg\textsuperscript{2+} ions (red) immobilized on ionized silanols
  (green). (\textbf{B}) Plot summarizes AFM data from
  (A). (\textbf{C}) AFM data (blue bars, $N=437$ sites) compared with
  kinetic (red) and thermodynamic (green) predictions for the
  percentage of right triangles bound at a given angle (within a
  9\degree\ macrostate) on a shape-matched binding site; Insets show
  models of the highest abundance microstate within a macrostate
  (purple triangles on green sites) and an AFM image of an example
  microstate.}
\vspace{-.1in}
\label{fig:fig2}
\end{figure*} 

The sequential combination of solution-phase self-assembly (SPSA) and
directed self-assembly (DSA) provides a general paradigm for the
synthesis of nanoscale devices and their large-scale integration with
control circuitry, microfluidics, or other conventionally-fabricated
structures. SPSA for the creation of sub-lithographic devices via
structural DNA nanotechnology \cite{Seeman2003} is relatively
mature. In particular, typical DNA origami \cite{rothemund2006folding}
allow up to 200 nanoscale components, including carbon nanotubes
\cite{maune2010origamiCNTfets, yan2013discreteCNTs,
  norton2013sitespecificCNT}, metal nanoparticles
\cite{AcunaNanoantenna2012, schreiber2014hierarchical}, fluorescent
molecules \cite{AcunaNanoantenna2012, schreiber2014hierarchical,
  gopinath2016starrynight}, quantum dots
\cite{schreiber2014hierarchical, QuantumDotLifetime2013liddle} and
conductive polymers \cite{conductive2015gothelf} to be simultaneously
juxtaposed at 3-5~nm resolution within a 100~nm$\times$70~nm DNA
rectangle. DSA uses topographic
\cite{liddle2004capillaryassembly,asbahi2015nanoparticlepositioning}
or chemical \cite{nealey2003classicblockcopolymerDSA,
  ross2014blockcopolymerpatterns, deyoreo2003virusesSPM,
  mirkin2005virusalignedDSA,
  mirkin2006carbonnanotubesDSA,zettl2006cntalign,
  ding2010tubeislandconnect, PearsonWoolleyOrigamiBlockCopolymer2011,
  Wind2013LinearOnGoldIslands,gothelf2016golddots,YanOrigamiOnGold2009,kershner2009originalplacement,
  penzo2011selective, gopinath2014optimizedplacement} patterning,
fields
\cite{dai2001electricCNTalign,correaduarte2006magneticalignment,mallouk2000metallicnanowires,Tanase2001magneticnanowires,Demirors2010electricdumbbells,nelson2009helicalmagneticswimmer,nelson2013MagneticSwimmerReview,KuzykYurke2008dielectrophoresis,toppari2015dielectrophoresisOrigami,mertig2016dielectrophoresisOrigami,DavisHowe2015antibodieselectricfield},
or flow
\cite{bensimon1995dnacombing,schwartz1998dnaflowaligned,li2013evaporativerediscovery,
  gothelf2015tubecombingalignment,dai2005germaniumwiresFluidAligned,yang2007alignedSilverWires,lieber2001orientednanowiresInP,li2015dnananowiresflow,guan2005pdmsstampcombing}
to control the higher order structure of molecules and
particles. Well-developed for continuous block copolymers films
\cite{nealey2003classicblockcopolymerDSA,
  ross2014blockcopolymerpatterns}, spherical nanoparticles
\cite{liddle2004capillaryassembly, asbahi2015nanoparticlepositioning},
and linear nanostructures \cite{mirkin2005virusalignedDSA,
  mirkin2006carbonnanotubesDSA, zettl2006cntalign,
  ding2010tubeislandconnect, PearsonWoolleyOrigamiBlockCopolymer2011,
  Wind2013LinearOnGoldIslands, gothelf2016golddots,
  dai2001electricCNTalign,correaduarte2006magneticalignment,mallouk2000metallicnanowires,Tanase2001magneticnanowires,Demirors2010electricdumbbells,nelson2009helicalmagneticswimmer,nelson2013MagneticSwimmerReview,KuzykYurke2008dielectrophoresis,toppari2015dielectrophoresisOrigami,mertig2016dielectrophoresisOrigami,
  bensimon1995dnacombing,schwartz1998dnaflowaligned,li2013evaporativerediscovery,
  gothelf2015tubecombingalignment,dai2005germaniumwiresFluidAligned,yang2007alignedSilverWires,lieber2001orientednanowiresInP,li2015dnananowiresflow,guan2005pdmsstampcombing},
DSA is less developed for origami-templated devices for which shape
and symmetry play an important role in device function and
integration.

Two challenges arise in the DSA of orgami-templated devices.  The
first is analogous to the problem of absolute orientation
\cite{horn1987absoluteorientation} (Fig.~1A) in computational
geometry: Given two Cartesian coordinate systems, what translation and
rotation can transform the first to the second?  Such transformations
are key in computer vision and robotics, where they can be used to
plan the motion of a virtual camera, or a robot arm. The physical
analog for DSA asks: How can an asymmetric device in solution be
positioned and aligned relative to a global reference frame in the
laboratory?  The second challenge is to achieve absolute orientation
for many devices at once, such that the position and alignment of each
device is arbitrary, i.e. independent of other devices (Fig.~1B). DNA
origami placement (DOP) \cite{kershner2009originalplacement, penzo2011selective,
  gopinath2014optimizedplacement} is a potential solution to both
challenges. In DOP the match between the overall shape of an origami
and lithographically patterned binding sites is used both to position
the origami in $x$ and $y$, and to control its in-plane rotation
$\theta$. The strength of DOP is that thousands of origami can be
oriented with high yield and fidelity: $\sim$95\% of sites have single
origami aligned within $\pm$10\degree\ of a desired $\theta$. The
weakness of DOP has been the exclusive use of equilaterial triangles:
an equilateral triangle can attach to its binding site in one of six
orientations (at any of three equivalent rotations, flipped right-side
up or up-side down). Thus DOP of equilateral triangles does not
achieve absolute orientation and its use is limited to devices with
compatible symmetry, e.g. point-like \cite{gopinath2016starrynight},
three-fold, or six-fold.

Consideration of fully asymmetric (C1~symmetric) devices, like bipolar
junction transistors, motivates the development of absolute and
arbitrary DSA (Fig.~1C), and clarifies conditions for which DOP of
high symmetry shapes (like equilateral triangles and rectangles) or
other DSA methods (fig.~S1) are insufficent. Were DOP of
rectangular origami used for the three-device circuit pictured, the
origami's symmetry would allow it to bind in four orientations
relative to each binding site: one (Fig.~1D) desired and three
(Fig.~1, E~to~G) undesired. Random binding at each site would result
in exponentially low yield: only $(0.25)^{3} = 1.6$\% of circuits
would have all three transistors in the desired orientation. Flow or
field alignment of induced dipoles would allow the same four
orientations. Field alignment of origami bearing fixed dipoles could
break in-plane rotational symmetry but would still allow two
orientations (Fig.~1D and F) related by a horizontal flip.  Further,
such purely global methods cannot simultaneously specify distinct
rotations or translations for multiple devices, and could not
fabricate the given circuit in a single step; arbitrary orientation
promises independent alignment of an unlimited number of devices in a
single step. Approaches which fix the ends of linear nanostructures on
metal bars or dots \cite{zettl2006cntalign, ding2010tubeislandconnect,
  Wind2013LinearOnGoldIslands}, or align them to chemical stripes
\cite{mirkin2005virusalignedDSA}, add arbitrary control of position
and in-plane rotation, but still cannot distinguish the orientations
in Fig.~1, D~to~G. Nor can methods which fix the corners of rectangles
\cite{gothelf2016golddots}. Here we show that absolute orientation can
be achieved by DOP with suitably asymmetric DNA origami shapes, and
demonstrate two applications in which absolute and arbitrary
orientation work together to optimize or integrate optical devices.

DOP can been performed on any planar substrate (e.g.
SiO\textsubscript{2}, quartz, silicon nitride [SiN] and diamond-like
carbon) whose surface can be differentiated into negatively-charged
binding sites (green features throughout paper) which bind
negatively-charged DNA origami strongly in the presence of bridging
Mg\textsuperscript{2+} ions, and a neutral background which binds
origami weakly (gray backgrounds). Here e-beam patterned binding
sites are made negative via silanols which are ionized at the pH (8.3)
of the origami binding buffer and the neutral background is a
trimethylsilyl monolayer, generated via silanization. DOP is a complex
adsorption process which involves both 3D diffusion to the surface,
and 2D diffusion of weakly bound origami on the
background. Observations of lateral jamming, binding of multiple
origami to a single site, and reorientation of origami already bound
to sites suggest that DOP is both nonequilibrium and non-Langmuir
\cite{gopinath2014optimizedplacement}.  Thus to simplify development
of absolute orientation, we separated the problem into two parts:
first, breaking up-down symmetry on unpatterned SiO\textsubscript{2}
(e.g. differentiating between the pair of orientations in Fig.~1, D
and E and the pair in Fig.~1, F and G) and second, breaking rotational
symmetry in the context of DOP (e.g. differentiating between Fig.~1D
and Fig.~1E).

\begin{figure*}[!hb]    
\centerline{ \includegraphics[width=\fonesize]{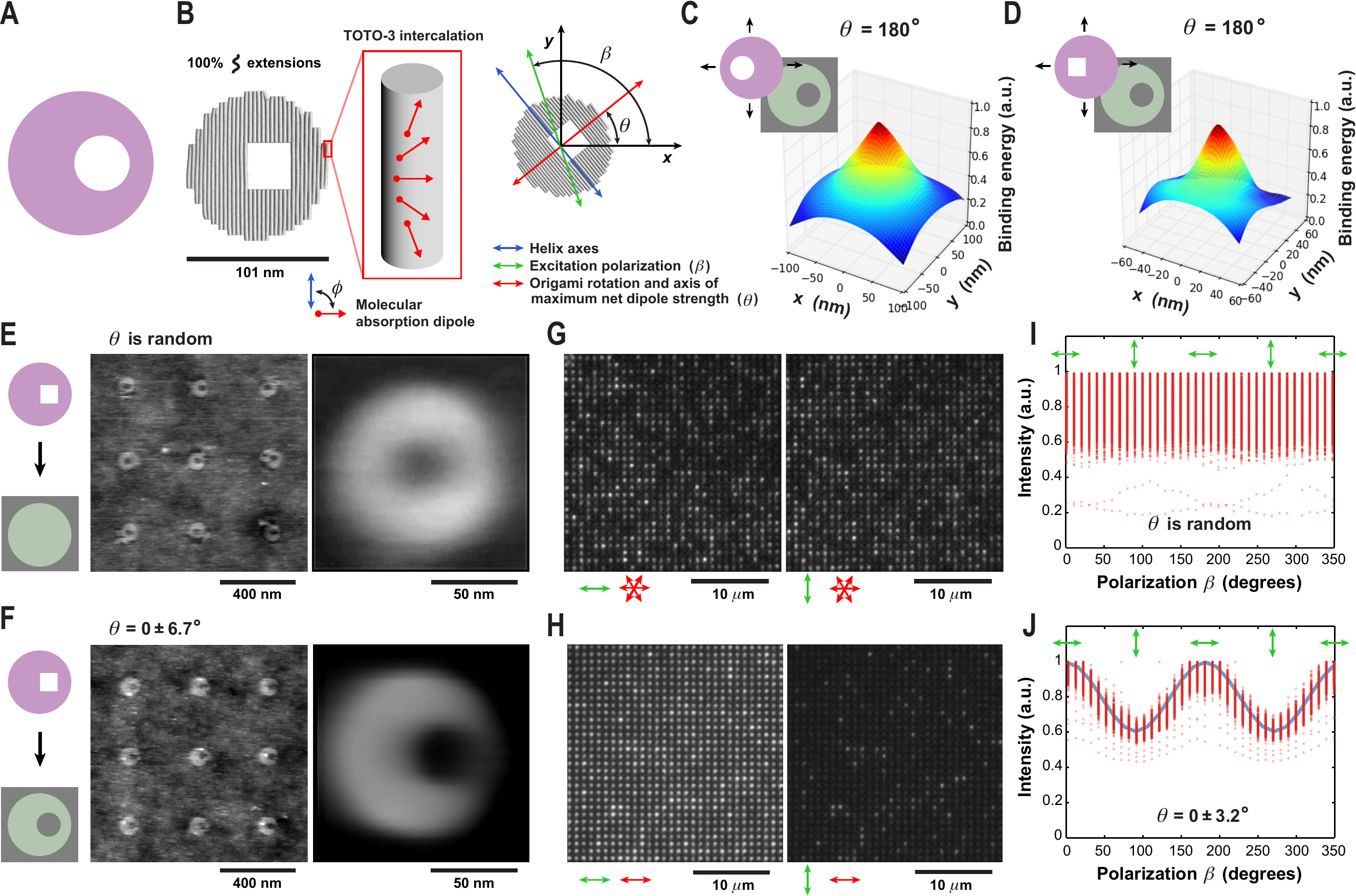} }  \vspace{-.1in}
\renewcommand{\baselinestretch}{\figcaptionbaselinestretch}
\caption{\bsf{Breaking rotational symmetry.} (\textbf{A}) Ideal ring
  with offset hole.  (\textbf{B}) DNA origami approximation of (A)
  comprised of 34 parallel helix axes (gray cylinders). Inset shows
  rotation of the fluorescent dye TOTO-3's absorption dipole along the
  length of a TOTO-3 intercalated helix. Coordinate system shows
  relationships between helix axes, excitation polarization ($\beta$)
  and origami rotation ($\theta$).  (\textbf{C}) Section of energy
  landscape for ideal shape (A) on binding site,
  $\theta$=180\degree. Colors run from high binding energy (red) to
  low (blue). (\textbf{D}) Same as (C), for experimental shape
  (B). (\textbf{E} and \textbf{F}) AFM and averaged AFM ($N$=600) of
  DOP on arrays of disk-shaped and shape-matched sites. (\textbf{G}
  and \textbf{H}) Fluorescence microscopy of TOTO-3 intercalated into
  DOP arrays on disk-shaped and shape-matched sites (ex. 642~nm;
  em. 660~nm). (\textbf{I} and \textbf{J}) Intensity (red dots) of
  $N$=600 sites in (G) and (H) as a function of excitation
  polarization $\beta$. Blue line, best fit.}
\vspace{-.1in}
\label{fig:fig3}
\end{figure*} 

The breaking of up-down symmetry was explored using asymmetric right
triangles (Fig.~2A). Synthesized via the SPSA of 200 short DNA staple
strands with a long scaffold strand, asymmetric right triangles have
left (orange) and right (purple) faces which are easily distinguished
by atomic force microscopy (AFM). Our idea was to make one side of the
origami non-sticky and hence bias binding, through the addition of
single-stranded (ssDNA) extensions to the 5$'$ ends of staples. To
control for geometric details of the right triangle design, and
isolate intrinsic bias which might arise from these details instead of
ssDNA extensions, two versions were created. In one version, the ends
of all staple strands and hence all nicks in the phosphate backbone
fell on the origami's right face; in the other, vice versa. Designed
to be flat via twist correction \cite{DietzCurvedOrigami2009},
extension-less right triangles of both types exhibited a weak
preference to bind unpatterned SiO\textsubscript{2} with their right
face up ($\sim$60:40 right:left, Fig.~2A); thus intrinsic bias was not
due to asymmetric flexibility caused by nick position. Bias has been
observed in curved single-sheet structures elsewhere
\cite{wei2013curvedSST,marchi2014toward} suggesting that residual
curvature due to imperfect twist correction of the right triangle
designs might be responsible for bias here. Strong bias (nearly 100\%)
was attained by adding 20~nt poly(T) ssDNA extensions to the ends of
all 200 staples; origami whose left face was extended bound left-face
up, and vice versa (Fig.~1B). Adding poly(A) ssDNA to make all
extensions double-stranded and rigid abolished the bias, supporting
the idea that on SiO\textsubscript{2} ssDNA extension create bias by
acting as entropic brushes which interfere with
DNA-SiO\textsubscript{2} binding. However, the symmetry-breaking
effect of ssDNA extensions on SiO\textsubscript{2} does not generalize
to other surfaces: on mica, where DNA-mica interactions are much
stronger than DNA-SiO\textsubscript{2} interactions for the same
Mg\textsuperscript{2+} concentration
\cite{gopinath2014optimizedplacement}, no bias was observed; on
graphene, where $\pi$-$\pi$ interactions between the unpaired bases
and graphene are attractive \cite{husale2010ssdna}, the bias inverted.


To break rotational symmetry, we began with the DOP of right-face
extended triangles (Fig.~2C), used the results to develop a model of
binding, and then used the model to design an origami shape which
achieved absolute orientation. AFM images of sites binding a single
right triangle (73\% of $N=$ 600 sites, fig.~S2) were analyzed, and
the angle $\theta$ between origami and binding site was measured to
the nearest multiple of 4.5\degree.  Only 34\% of origami bound with
the desired alignment ($\theta = 0$\degree), too few for reliable
absolute orientation. We next asked whether the distribution of states
better fit a kinetic or equilibrium model, under the assumption that
the binding energy of a given state is linearly proportional to the
area of overlap between the origami and binding site; $\theta = 0$,
with its total overlap of origami and binding site, has the highest
possible binding energy. The state space was discretized in both $x$
and $y$ (1~nm increments), and $\theta$ (1\degree\ increments),
encompassing more than 19 million states with positive overlap. For
kinetic predictions (Fig.~2C, red), we performed steepest ascent hill
climbing using all possible states as initial configurations, and
found that (neglecting variations in $x$ and $y$) the state space had
three basins of attraction whose maxima ($\theta = 0$, $\pm
150$\degree) corresponded with the three most common experimental
states (Fig.~2C, blue). Kinetic abundances predicted by measuring and
normalizing basin volumes overestimated experimental abundances with
relatively small factors (from 1.1$\times$ for $\theta = 0$\degree\ to
2.6$\times$ for $150$\degree). Small changes to details of the model
(Fig.~S3) predicted the existence but not quantitative abundance of
minority states (e.g. $\theta = -122$ or 90\degree). For thermodynamic
predictions (Fig.~2C, green), we calculated expected equilibrium
abundances from the partition function, using an energy per unit area
overlap derived by constraining the abundance at $\theta =
0\pm4$\degree\ to match experiment; thermodynamic abundances
underestimated experimental abundances with large factors (from
5.5$\times$ for $\theta = 150$\degree\ to 7.3$\times$ for
$-150$\degree).  Our data are thus most consistent with a strongly
kinetically trapped regime in which origami enter the state space at
random (when they collide with a binding site) and simply proceed to a
local maxima (fig.~S4A) in binding energy.

\begin{figure*}[!ht]    
\centerline{ \includegraphics[width=\fonesize]{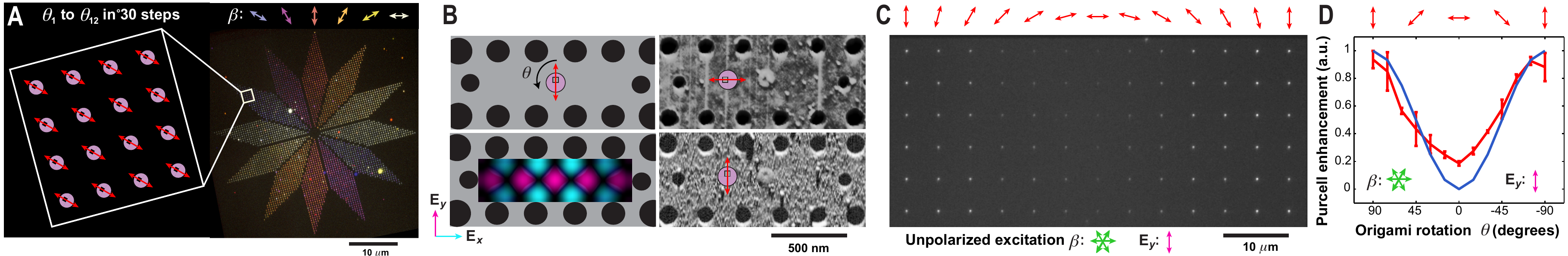} }  \vspace{-.1in}
\renewcommand{\baselinestretch}{\figcaptionbaselinestretch}
\caption{\bsf{Applications of orientation.} ((\textbf{A}) 2D polarimeter
  composed of 3,456 origami divided into 12 rhomboidal arrays, with
  $\theta$ graduated in 30\degree steps. Combination of 36
  fluorescence images (fig.~S12) colored according to $\beta$ enables
  six polarizations to be distinguished by eye. (\textbf{B}) Schema,
  simulation, and AFM for coupling between TOTO-3 emitters and PCCs as
  a function of origami rotation $\theta$. $E_\mathrm{y}$ (purple) and
  $E_\mathrm{x}$ (blue) label polarization for FDTD simulation
  \cite{gopinath2016starrynight} of the electric field. Red axes show
  polarization of peak TOTO-3 emission. (\textbf{C}) Fluorescence of a
  PCC array with varying $\theta$, excited by unpolarized light
  $\beta$. Maximum coupling is observed when origami align TOTO-3
  emission dipoles with $E_\mathrm{y}$. (\textbf{D}) Data (red) from
  (C) and simulation (blue). Error bars, $\pm1$SD for $N$=6.}
\vspace{-.1in}
\label{fig:fig4}
\end{figure*} 

The strong kinetic trapping exhibited by DOP constrains the energy
landscapes which can robustly break rotational symmetry: the volume of
a single basin of attraction must comprise most of the state space; in
the best case the landscape will have a unique global maximum. Exact
analysis \cite{bohringer2004offsetringdesign} and general yet simple
geometric arguments \cite{gopinath2016alignment} have shown existence
of a unique global maximum for a disk with an offset hole (Fig.~3A), a
shape we call a `small moon'.  Experiments with millimeter-scale
models on hydrophobic binding sites
\cite{bohringer2004capillaryshapematching} confirm that small moons
translate and rotate to a unique orientation from initial
configurations created by hand using tweezers. Here, we approximated
the small moon shape by a DNA origami (Fig.~3B, fig.~S5A) with an
offset square hole (circumscribed by the ideal hole). Exact
mathematical analysis of the energy landscape of the approximate small
moon was hindered by its complex jagged outline, so we discretized the
landscape as above. Like its idealized counterpart the DNA origami
small moon has a unique global maximum in its energy landscape,
although the square-shaped hole slightly flattens the landscape in
some regions (compare Figs.~3C and D, figs.~S4B and C). DOP of small
moon origami with ssDNA extensions to break up-down symmetry
(fig.~S5D) was performed on both disk-shaped control sites (Fig.~3E
and fig.~S6) and shape-matched sites (Fig.~3F and fig.~S7).  The
average of 498 AFM images of control sites with single origami (83\%
of 600 total sites) gave an annular shape indicating random
orientation; the average of 592 images on shape-matched sites (98.7\%
of 600 total sites) reconstruct the small moon shape, confirming
unique alignment.

By fitting the small moon shape to AFM of small moon origami on
shape-matched sites, we found that alignment varied by
$\pm$6.7\degree\ ($\pm$1 SD). This variability includes both real
variability due to fabrication error or imperfect assembly, and
spurious variability due to the fitting of a model shape to poorly
resolved origami; the latter error is difficult to estimate. To get a
better estimate of alignment precision, we imaged small moons
intercalated post-DOP with the fluorescent dye TOTO-3 (Figs. 3G to J,
figs.~S8 and S9). For 600-site arrays of small moons on disk-shaped
control and shape-matched sites, we measured emission intensity for
excitation polarization $\beta$ in 10\degree\ steps (sampling each
$\beta$ twice by rotating the stage from 0\degree\ to 350\degree) and
fit the emission to derive distributions for the origami orientation
$\theta$. The reported angle between the molecular absorption dipole
of TOTO-3 analogs and the DNA helix axis ($\phi$) ranges from
61\degree\ to 90\degree\ \cite{spielmann1995TOTO1nmr90degrees,
  schins1999toto1at61egrees, bennink1999yoyo1anisotropy69degrees,
  persson2009yoyo1anisotropy86degrees}, but the exact angle is
unimportant for measuring variability: it is close enough to
90\degree\ that averaging over multiple dyes (intercalated at varying
rotations due to twist, Fig.~3B) results in a strongly anisotropic net
dipole strength in the plane of the origami. Consequently, emission
peaks for $\beta$ perpendicular to the helix axes
\cite{persson2009yoyo1anisotropy86degrees}, coincident with
$\theta$. The strength of a molecular dipole $\bm{\vec{\mu}}$ excited
by an electric field $\vec{E}$ along the direction of unit vector
$\bm{\hat{e}} = \vec{E}/|\vec{E}|$ is $D(\vec{E}) =|\bm{\vec{\mu}}
\boldsymbol{\cdot} \bm{\hat{e}}|^2 = |\bm{\vec{\mu}}|^2
\cos^2(\beta-\theta)$ where $\beta$ is the polarization of $\vec{E}$,
and $\theta$ the in-plane dipole angle. According to the dipole
approximation \cite{ha1999polarization,
  novotny2000singlemoleculepolarization}, emission is proportional to
absorption, which is proportional to $|\vec{E}|^2 D(\vec{E})$. Thus
experimental intensity can be fit to $I_{o} \cos^{2}(\beta-\theta) +
c$ where $I_o$ is the maximum emission, and $c$ is the background
(camera noise, reflection). Emission from a collection of $n$
molecular dipoles $\bm{\vec{\mu}}_k$ bound to an origami is
proportional to $|\vec{E}|^2 D_{\mathrm{net}}$, where the net dipole
strength\footnote{Note that the strength of the net dipole moment is
  not the same as the net dipole strength. Consider equal and opposite
  dipoles $\bm{\vec{\mu}}_{\shortuparrow} = -
  \bm{\vec{\mu}}_{\shortdownarrow}$ intercalated 180\degree\ from each
  other around the helix. They cancel to yield zero net dipole moment
  but contribute equally to the net dipole strength, and hence
  emission under $\vec{E}$.}  is given by $D_{\mathrm{net}}(\vec{E}) =
\sum_{k=1}^{n} |\bm{\vec{\mu}}_k \boldsymbol{\cdot}
\bm{\hat{e}}|^2$. Thus the experimental intensity of $n$ molecular
dipoles with an anisotropic net in-plane dipole strength can be fit to
the $\cos^{2}$ expression above: if $\vec{E}_{\parallel}$ and $\theta$
are defined to lie along the direction of maximum net dipole strength,
then $I_o$ is proportional to the difference
$D_{\mathrm{net}}(\vec{E}_{\parallel}) -
D_{\mathrm{net}}(\vec{E}_{\perp})$ and $c$ is the background plus a
contribution proportional to $D_{\mathrm{net}}(\vec{E}_{\perp})$, from
the direction of smallest net dipole strength. Emission from control
sites (Fig.~3I, fig.~S10A and B) individually fit this expression but
individual $\theta$ were uniformly distributed (fig.~S10C), both
confirming random origami orientation and ruling out polarization
anisotropy in our setup. As expected, aggregate data could not be
fit. In contrast, aggregate data for shape-matched sites (Fig.~3J) fit
$\theta = $ 0\degree\ and fits to individual sites (fig.~S10D) vary by
$\pm$3.2\degree, our best estimate of alignment precision.

TOTO-3 intercalation of small moons further enabled us to demonstrate arbitrary orientation, prototype the large-scale integration of orientation-dependent devices, and explore variables which can affect the quality of polarization-based devices. However, even when the $\theta$ and $\beta$ are orthogonal to each other there is a small, but reproducible, excitation of the light emitters which we refer to as the bleed-through of the system. We quantified bleed-through for the data in Fig. 3J; after background subtraction we found that emission from origami perpendicular to $\beta$ was 30\% of that from origami parallel to β. We quantified
bleed-through for the data in Fig.~3J; after background subtraction we
found that emission from origami perpendicular to $\beta$ was \fx{30\%} of
that from origami parallel to $\beta$. In interpreting the source of
bleed-through, we consider only the effect of dye alignment
and neglect small polarization mixing effects of high numerical
aperture on excitation polarization \cite{ha1999polarization}.
In an ideal device, all dye molecules would align perfectly with
$\vec{E}_{\parallel}$: $D_{\mathrm{net}}(\vec{E}_{\perp})$ and hence
bleed-through would be zero. $D_{\mathrm{net}}(\vec{E}_{\perp})$
combines contributions from both placement variability in $\theta$
with incoherence of dye angle relative to the origami.  The
contribution from placement variability is small, as bleed-through
would be only 0.3\% were the $\pm$3.2\degree\ variability the only
source; $\pm$39\degree\ variability would be required to explain
30\% bleed-through. The contribution from incoherent dye alignment
within an origami is itself complex: it combines the deterministic
rotation of $\phi$ by DNA twist, random wobble
\cite{BackerMoerner2016singlemoleculepolarization,
  brasselet2016singlemoleculepolarization} from rotational diffusion
(reduced here by intercalation and drying), potential alternative
binding modes \cite{milanovich1996twobiningmodesTOTO3}, and
significant ($\sim$10.6\degree, fig.~S11) back-and-forth bending of
each helix axis in a DNA origami \cite{rothemund2006folding}. Here
we explain bleed-through simply by a combination of $\phi$ and helix
bending, which are the most relevant variables for devices based on
intercalators. Attributing all bleed-through to the dipole-helix angle
yields \fx{$\phi = $ 69\degree}\ and adding helix bending increases our
estimate of \fx{$\phi$ to 70}\degree; both are consistent with $\phi$
previously measured for TOTO-3 analogs. As with the addition of helix
bending, adding other sources of dye alignment incoherence or
excitation polarization mixing to the model would increase our
estimate of $\phi$; thus given our data, \fx{69\degree}\ is a
lower bound for $\phi$. On the other hand, even if $\phi =
$ 90\degree\ were achieved and all other sources of alignment
incoherence removed, helix bending would still cause $\sim$3.5\%
bleed-through, an unavoidable consequence of randomly intercalating
dyes binding to both $+10.6$\degree\ and $-10.6$\degree\ bent helices.
Devices with better-defined alignment relative to DNA origami, such as
gold rods \cite{pal2011rodsonorigami,kuzyk2014rodsalign} or single
site-specific rigidly-linked chromophores
\cite{vybornyi2014porphyrinDNAcovalent}, would exhibit much stronger
polarization effects, limited only by the placement variability
(i.e. 0.3\% bleed-through might be attained).

Despite the limitations of intercalating dyes, Fig.~4A shows that
arbitrary orientation can integrate 3,456 TOTO-3 labelled small moons
with 12 different $\theta$ into a microscopic fluorescent polarimeter,
a 100~\um\ device which glows most strongly along the polarization
axis of incident light.  Microscopic polarimeters constructed using
plasmonic antennas have been created in the near-IR
\cite{brongersma2012polarimeter}, and arrays of oriented gold rods
have been used for metasurface polarimeters at telecommunication
wavelengths \cite{capasso2016metasurfacepolarimeter}; the goal of such
on-chip instruments is to replace multiple bulky and expensive optical
components and to make {\em in situ} measurements possible, within
devices or transmission lines. Since our polarimeter reports
polarization directly, it could be fabricated on microscope slides and
used {\em in situ} to aid polarized fluorescence microscopy
\cite{oldenbourg2016pnaspolarizationmicroscopy}: to align excitation
polarization grossly by eye without requiring analyzers, to check for
polarization bias, or as a calibration standard for fluorescence
anisotropy of biomolecules. Operating wavelength could be tuned via
intercalation of different dyes (e.g. YOYO-1, 491~nm excitation;
TOTO-1, 514~nm; YOYO-1, 612~nm; TOTO-3, 642~nm), or made broadband by
using a mixture. Based on the $\pm$3.2\degree\ variability we observe,
fitting the orientation of 3,456 origami would allow the angle between
excitation polarization and surface features to be measured with a
precision of 0.05\degree\ (SEM).  Our polarimeter is unable to measure
$z$-polarization, but DOP of 3D origami could add this capability. And
while our polarimeter is not a metasurface
\cite{capasso2014metasurfaces}, it provides a roadmap for how DOP
could push metal-rod metasurfaces from the near-IR, where the rods are
fabricated lithographically, to the visible, via oriented arrays of
smaller colloidal gold rods
\cite{pal2011rodsonorigami,kuzyk2014rodsalign}.

 Hybrid nanophotonic devices \cite{benson2011assembly} combine light
 emitters or scatterers with microfabricated optical resonators to
 obtain devices ranging from biosensors
 \cite{ArmaniFraserFlaganVahala2007LabelFreeDetectionInMicrocavity} to
 light sources for on-chip quantum information processing
 \cite{hennessy2007stochasticQD}. The performance (e.g. sensitivity of
 a detector, or intensity of a light source) of such devices hinges on
 the strength of the coupling between the emitter and resonator. In
 particular, emission intensity is proportional to the cavity Purcell
 enhancement $F_{\mathrm{cav}} \propto |\bm{\vec{\mu}}
 \boldsymbol{\cdot} \vec{E}(\vec{r})|^2$, which is typically a
 sensitive function of the position of the emitter $\vec{r}$, and the
 orientation of the emission dipole $\bm{\vec{\mu}}$ relative to the
 cavity electric field $\vec{E}$ \cite{riedrich2014deterministic}. To
 maximize coupling, the emitter should be positioned in a peak of a
 resonant mode, with $\bm{\vec{\mu}}$ aligned to the polarization of
 $\vec{E}$ at $\vec{r}$.  Fabricating resonators with simultaneously
 positioned and aligned emitters has been a difficult challenge
 \cite{maitre2014measureandselectorientation}. Most approaches for
 positioning involve randomly growing or depositing emitters on a
 surface, selecting emitters with microscopy, and tediously
 fabricating resonators around them \cite{hennessy2007stochasticQD,
   tartakovskii2013singlephotonstochasticQDonPCC,
   riedrich2014deterministic,
   sapienza2015singlephotonstochasticQDonbragg}.  Some emitters can be
 grown at predetermined sites within resonators
 \cite{LyasotaKapon2015lithographicallydirectedQDs}, but in general,
 deterministic approaches for positioning emitters rely on scanning
 probe microscopy
 \cite{barth2010nanoassembled,englund2010deterministic1DdiamondNV}.
 Neither ``select and post-process'' nor scanning probe approaches can
 scale to large numbers of devices, or provide deterministic
 alignment. Conversely, methods for achieving deterministic alignment
 of molecular or vacancy-based emitters \cite{sandoghdar2004terrylene,
   ToninelliSandoghdar2010alignedmoleculesNearIR,
   Polisseni2016aligneddibenzoterryleneemitters,
   LesikJacques2014alignedNVcenters,
   MichlWrachtrup2014alignedNVcenters} do not address positioning.
 Previously \cite{gopinath2016starrynight}, we used DOP to achieve the
 large-scale positioning of molecular emitters within L3 photonic
 crystal cavities (PCCs); TOTO-3 intercalated small moons allowed us
 to extend that work to control the alignment $\theta$ of
 $\bm{\vec{\mu}}$ in the cavity (Fig.~4, B~to~D). To optimize emission
 from the PCCs, we created a 13$\times$6 array of identical resonators
 (fig.~S13 and S14) with small moons positioned in the center of a
 $y$-polarized peak in $\vec{E}$, and varied $\theta$ in 13 steps from
 90\degree\ to -90\degree across the width of the array. Emission
 intensity roughly followed the expected $\cos^{2}(\theta)$
 relationship, and a 4.5-fold increase was observed for $\theta$ which
 maximally align TOTO-3 dipoles with $\vec{E_y}$.  Potential reasons
 for disagreement between experimental intensity at 0\degree\ with
 FDTD simulation of a single dipole are similar to those for
 bleed-through above: TOTO-3 dyes are spread out over the 100~nm
 diameter disk of the small moons rather than in the exact center of
 the cavity, $\phi \neq 90$ contributes to a net dipole strength
 parallel to $\vec{E_x}$, and alignment error.  Beyond
 emitter-in-cavity devices, our ability to simultaneously position and
 orient molecular and nanoparticle components should find wide use in
 nanophotonics. The collective behavior of multiple emitter systems is
 highly sensitive to inter-emitter distance and relative dipole
 orientation, suggesting that our technique will be ideal for studying
 and engineering fundamental phenomena such as superradiance
 \cite{LimBardeen2004TetraceneSuperradianceInThinFilms}, and other
 coherence effects
 \cite{HettichSandoghdar2002coupledalignedMolecules}. Positioning and
 orientation of molecular emitters within optical nanoantennas will
 allow antenna performance to be optimized
 \cite{Novotny2011PhysicsTodayNearFieldOptics}; similar control over
 metal nanoparticle dipoles will enable optical nanocircuit elements
 to be programmed with series, parallel or intermediate behavior
 \cite{SalandrinoEngheta2007OpticalNanocircuitsPhysical,
   AluEngheta2008OpticalNanoantennasNanocircuits}.

We have engineered the energy landscape of DNA origami shapes on
binding sites to realize absolute and arbitrary orientation, enabling
DSA to independently specify all degrees of freedom and thus break all
translational and rotational symmetries for arbitrary numbers of
C1-symmetric molecular devices. Perhaps surpisingly, we achieved this
by combining broken up-down symmetry with a mirror symmetric (D1,
bilateral) shape---the small moon; a fully asymmetric (C1) shape was
neither necessary nor sufficient---the C1-symmetric right triangle
suffered from kinetic trapping.\footnote{A system with multiple local
  maxima and a single global maximum could break rotational symmetry
  in the limit of slow annealing to zero temperature. We have yet to
  find a practical way to anneal DOP, but a combination of heat and
  monovalent cations has been used to mobilize and crystallize origami
  kinetically trapped on mica \cite{woo2014cation}.}  Yet the devices
we have presented do not demonstrate the full power of the small
moons---the two-fold degeneracy of transition dipoles means that D2
symmetric shapes, e.g. an elongate rectangle or oval, could have been
used. No isolated optical device, or coupled array of optical devices
yet designed seem to require full symmetry-breaking: 2D chiral
scatterers \cite{decker2007ChiralMetasurfaces} (C4) require up-down
symmetry to be broken but not rotational, U-shaped resonators (D1) for
certain nonlinear metasurface holograms
\cite{ye2016MonlinearMetasurfaceHolographyCD} require that complete
rotational symmetry be broken but not up-down. Within electronics, no
molecular device with the C1 symmetry of a bipolar junction transistor
has been achieved: molecular diodes
\cite{ratner1974rectifier,ratner2015newrectifier} (D1) can tolerate
flips about their mirror plane and crossed-CNT FETs
\cite{maune2010origamiCNTfets} (D2) can tolerate two flips and
180\degree\ rotation. On the other hand, proposed planar optical and
electronic circuits \cite{wong2013CNTcomputer} of even just a few
symmetric components can almost invariably take advantage of absolute
and arbitrary orientation to avoid tortuous paths for interconnect. In
part, applications for DSA of molecular components have been
constrained by what has been possible. Now that molecular orientation
can be controlled, we anticipate that new asymmetric devices and
architectures will be explored.

\thispagestyle{empty}
\svsp

{\bf Acknowledgments} {\small We acknowledge funding from Office of
  Naval Research Award N000141410702, U.S. National Science Foundation
  grant Nos. 1636364 and 1317694 (Expedition in Computing, Molecular
  Programming Project, {\tt \footnotesize
    http://molecular-programming.org}), Air Force Office of Scientific
  Research FA9550-16-1-0019 (A.M), the Natural Sciences and
  Engineering Research Council of Canada (D.K.), a Banting Fellowship
  (C.T.), the Orr Family Foundation, and the Abedin
  Institute. Fabrication was done at Caltech's Kavli Nanoscience
  Institute. 
  }

{\bf Corresponding authors.} Email: ashwing@caltech.edu (A.G.) or
pwkr@dna.caltech.edu (P.W.K.R).


{\bf Author Contributions} 
A.G. and P.W.K.R. conceived the project. A.G. performed origami synthesis, nanofabrication, AFM,
SEM, and fluorescence microscopy. C.T and D.K formalized proof for the deathstar origami design. A.G. and C.T. wrote the simulation code for surface reorientation model. All authors contributed to data interpretation and manuscript preparation.

\svsp {\bf Supplementary Materials} Materials and Methods, Figs.~S1 to
S14, References ({\em XX--YY}), DNA sequences, Design
Files, and Simulation Software.

\vspace{-0.1in}

\nocite{origamiPEGpurification2014}
\nocite{hogberg2015OrigamiPurification} 
\nocite{gopinath2014optimizedplacement}
\nocite{gopinath2016starrynight}
\nocite{noy2007handbook} 
\nocite{hogberg2012intercalatordistortion,choi2017intercalatororigamiswitching}
\nocite{ke2012intercalatingunderwinding}

\nocite{liddle2004capillaryassembly, deyoreo2003virusesSPM,
    asbahi2015nanoparticlepositioning}
\nocite{bensimon1995dnacombing,
    schwartz1998dnaflowaligned,li2013evaporativerediscovery,
    gothelf2015tubecombingalignment}
\nocite{dai2005germaniumwiresFluidAligned,yang2007alignedSilverWires}
\nocite{lieber2001orientednanowiresInP}
\nocite{li2015dnananowiresflow}
\nocite{lieber2001orientednanowiresInP,
    guan2005pdmsstampcombing}
\nocite{Wang2009niceNanowireDSAreview}
\nocite{dai2001electricCNTalign, correaduarte2006magneticalignment}
\nocite{mallouk2000metallicnanowires,
    Tanase2001magneticnanowires} 
\nocite{Demirors2010electricdumbbells}
\nocite{DavisHowe2015antibodieselectricfield}
\nocite{nelson2009helicalmagneticswimmer,nelson2013MagneticSwimmerReview}
\nocite{KuzykYurke2008dielectrophoresis,
    toppari2015dielectrophoresisOrigami,mertig2016dielectrophoresisOrigami}
\nocite{zettl2006cntalign}  
\nocite{mirkin2005virusalignedDSA}
\nocite{mirkin2006carbonnanotubesDSA} 
\nocite{ding2010tubeislandconnect,Wind2013LinearOnGoldIslands}
\nocite{PearsonWoolleyOrigamiBlockCopolymer2011} 
\nocite{gothelf2016golddots}
\nocite{kershner2009originalplacement, penzo2011selective,
    gopinath2014optimizedplacement}
\nocite{arbona2012interhelixGapCrossovers}

\renewcommand\refname{References and Notes}
\renewcommand{\bibfont}{\normalfont\small} 
\bibliography{AbsoluteOrientation}

\begin{thebibliography}{100}

\bibitem{Seeman2003}
N.~C. Seeman.
\newblock {DNA} in a material world.
\newblock {\em Nature\/}, 421:427--431, 2003.

\bibitem{rothemund2006folding}
P.~W.~K. Rothemund.
\newblock Folding {DNA} to create nanoscale shapes and patterns.
\newblock {\em Nature\/}, 440(7082):297--302, 2006.

\bibitem{maune2010origamiCNTfets}
H.~T. Maune, S.-P. Han, R.~D. Barish, M.~Bockrath, W.~A. Goddard~III, et~al.
\newblock {S}elf-assembly of carbon nanotubes into two-dimensional geometries
  using {DNA} origami templates.
\newblock {\em Nat. Nanotechnol.\/}, 5(1):61--66, 2010.

\bibitem{yan2013discreteCNTs}
Zhao Zhao, Yan Liu, and Hao Yan.
\newblock {DNA} origami templated self-assembly of discrete length single wall
  carbon nanotubes.
\newblock {\em Org. Biomol. Chem.\/}, 11:596--598, 2013.

\bibitem{norton2013sitespecificCNT}
Anshuman Mangalum, Masudur Rahman, and Michael~L. Norton.
\newblock Site-specific immobilization of single-walled carbon nanotubes onto
  single and one-dimensional {DNA} origami.
\newblock {\em Journal of the American Chemical Society\/}, 135(7):2451--2454,
  2013.

\bibitem{AcunaNanoantenna2012}
G.~P. Acuna, F.~M. M{\"o}ller, P.~Holzmeister, S.~Beater, B.~Lalkens, et~al.
\newblock Fluorescence enhancement at docking sites of {DNA}-directed
  self-assembled nanoantennas.
\newblock {\em Science\/}, 338(6106):506--510, 2012.

\bibitem{schreiber2014hierarchical}
R.~Schreiber, J.~Do, E.-M. Roller, T.~Zhang, V.~J. Sch{\"u}ller, et~al.
\newblock Hierarchical assembly of metal nanoparticles, quantum dots and
  organic dyes using {DNA} origami scaffolds.
\newblock {\em Nature Nanotechnology\/}, 9(1):74--78, 2014.

\bibitem{gopinath2016starrynight}
Ashwin Gopinath, Evan Miyazono, Andrei Faraon, and Paul W.~K. Rothemund.
\newblock Engineering and mapping nanocavity emission via precision placement
  of {DNA} origami.
\newblock {\em Nature\/}, 535:401--405, 2016.

\bibitem{QuantumDotLifetime2013liddle}
S.~H. Ko, K.~Du, and J.~A. Liddle.
\newblock Quantum-dot fluorescence lifetime engineering with {DNA} origami
  constructs.
\newblock {\em Angewandte Chemie International Edition\/}, 52(4):1193--1197,
  2013.

\bibitem{conductive2015gothelf}
Jakob~Bach Knudsen, Lei Liu, Anne Louise~Bank Kodal, Mikael Madsen, Qiang Li,
  et~al.
\newblock Routing of individual polymers in designed patterns.
\newblock {\em Nature Nanotechnology\/}, 10:892--898, 2015.

\bibitem{liddle2004capillaryassembly}
J.~Alexander Liddle, Yi~Cui, and Paul Alivisatos.
\newblock Lithographically directed self-assembly of nanostructures.
\newblock {\em J. Vac. Sci. Technol. B\/}, 22:3409--3414, 2004.

\bibitem{asbahi2015nanoparticlepositioning}
Mohamed Asbahi, Shafigh Mehraeen, Fuke Wang, Nikolai Yakovlev, Karen S.~L.
  Chong, et~al.
\newblock Large area directed self-assembly of sub-10 nm particles with single
  particle positioning resolution.
\newblock {\em Nano Letters\/}, 15(9):6066--6070, 2015.

\bibitem{nealey2003classicblockcopolymerDSA}
Sang {Ouk Kim}, Harun~H. Solak, Mark~P. Stoykovich, Nicola~J. Ferrier, Juan~J.
  de~Pablo, et~al.
\newblock Epitaxial self-assembly of block copolymers on lithographically
  defined nanopatterned substrates.
\newblock {\em Nature\/}, 424:411--414, 2003.

\bibitem{ross2014blockcopolymerpatterns}
Jae-Byum Chang, Hong~Kyoon Choi, Adam~F. Hannon, Alfredo Alexander-Katz,
  Caroline~A. Ross, et~al.
\newblock Design rules for self-assembled block copolymer patterns using tiled
  templates.
\newblock {\em Nature Communications\/}, 5:3305, 2014.

\bibitem{deyoreo2003virusesSPM}
Chin~Li Cheung, Julio~A. Camarero, Bruce~W. Woods, Tianwei Lin, John~E.
  Johnson, et~al.
\newblock Fabrication of assembled virus nanostructures on templates of
  chemoselective linkers formed by scanning probe nanolithography.
\newblock {\em Journal of the American Chemical Society\/}, 125(23):6848--6849,
  2003.

\bibitem{mirkin2005virusalignedDSA}
Rafael~A. Vega, Daniel Maspoch, Khalid Salaita, and Chad~A. Mirkin.
\newblock Nanoarrays of single virus particles.
\newblock {\em Angewandte Chemie International Edition\/}, 44(37):6013--6015,
  2005.

\bibitem{mirkin2006carbonnanotubesDSA}
Yuhuang Wang, Daniel Maspoch, Shengli Zou, George~C. Schatz, Richard~E.
  Smalley, et~al.
\newblock Controlling the shape, orientation, and linkage of carbon nanotube
  features with nano affinity templates.
\newblock {\em Proceedings of the National Academy of Sciences\/},
  103:2026--31, 2006.

\bibitem{zettl2006cntalign}
T.~D. Yuzvinsky, A.~M. Fennimore, A.~Kis, and A.~Zettl.
\newblock Controlled placement of highly aligned carbon nanotubes for the
  manufacture of arrays of nanoscale torsional actuators.
\newblock {\em Nanotechnology\/}, 17(2):434, 2006.

\bibitem{ding2010tubeislandconnect}
Baoquan Ding, Hao Wu, Wei Xu, Zhao Zhao, Yan Liu, et~al.
\newblock Interconnecting gold islands with {DNA} origami nanotubes.
\newblock {\em Nano Letters\/}, 10(12):5065--5069, 12 2010.

\bibitem{PearsonWoolleyOrigamiBlockCopolymer2011}
Anthony~C. Pearson, Elisabeth Pound, Adam~T. Woolley, Matthew~R. Linford,
  John~N. Harb, et~al.
\newblock Chemical alignment of {DNA} origami to block copolymer patterned
  arrays of 5 nm gold nanoparticles.
\newblock {\em Nano Lett.\/}, 11(5):1981--1987, 2011.

\bibitem{Wind2013LinearOnGoldIslands}
Risheng Wang, Matteo Palma, Erika Penzo, and Shalom~J. Wind.
\newblock Lithographically directed assembly of one-dimensional {DNA}
  nanostructures via bivalent binding interactions.
\newblock {\em Nano Research\/}, 6:409--417, 2013.

\bibitem{gothelf2016golddots}
Piero Morales, Liqian Wang, Abhichart Krissanaprasit, Claudia Dalmastri, Mario
  Caruso, et~al.
\newblock Suspending {DNA} origami between four gold nanodots.
\newblock {\em Small\/}, 12(2):169--173, 2016.

\bibitem{YanOrigamiOnGold2009}
Aren~E. Gerdon, Seung~Soo Oh, Kuangwen Hsieh, Yonggang Ke, Hao Yan, et~al.
\newblock {C}ontrolled {D}elivery of {DNA} {O}rigami on {P}atterned {S}urfaces.
\newblock {\em Small\/}, 5(17):1942--1946, 2009.

\bibitem{kershner2009originalplacement}
R.~J. Kershner, L.~D. Bozano, C.~M. Micheel, A.~M. Hung, A.~R. Fornof, et~al.
\newblock {P}lacement and orientation of individual {DNA} shapes on
  lithographically patterned surfaces.
\newblock {\em Nat. Nanotechnol.\/}, 4(9):557--561, 2009.

\bibitem{penzo2011selective}
Erika Penzo, Risheng Wang, Matteo Palma, and Shalom~J. Wind.
\newblock {S}elective {P}lacement of {DNA} {O}rigami on {S}ubstrates
  {P}atterned by {N}anoimprint {L}ithography.
\newblock {\em J. Vac. Sci. Technol. B\/}, 29(6):06F205, 2011.

\bibitem{gopinath2014optimizedplacement}
A.~Gopinath and P.~W.~K. Rothemund.
\newblock Optimized assembly and covalent coupling of single-molecule {DNA}
  origami nanoarrays.
\newblock {\em ACS Nano\/}, 8(12):12030--12040, 2014.

\bibitem{dai2001electricCNTalign}
Yuegang Zhang, Aileen Chang, Jien Cao, Qian Wang, Woong Kim, et~al.
\newblock Electric-field-directed growth of aligned single-walled carbon
  nanotubes.
\newblock {\em Applied Physics Letters\/}, 79(19):3155--3157, 2001.

\bibitem{correaduarte2006magneticalignment}
Miguel~A. Correa-Duarte, Marek Grzelczak, Ver\'{o}nica Salgueiri{\~n}o-Maceira,
  Michael Giersig, Luis~M. Liz-Marz\'{a}n, et~al.
\newblock Alignment of carbon nanotubes under low magnetic fields through
  attachment of magnetic nanoparticles.
\newblock {\em The Journal of Physical Chemistry B\/}, 109(41):19060--19063,
  2005.

\bibitem{mallouk2000metallicnanowires}
Peter~A. Smith, Christopher~D. Nordquist, Thomas~N. Jackson, Theresa~S. Mayer,
  Benjamin~R. Martin, et~al.
\newblock Electric-field assisted assembly and alignment of metallic nanowires.
\newblock {\em Applied Physics Letters\/}, 77(9):1399--1401, 2000.

\bibitem{Tanase2001magneticnanowires}
Monica Tanase, Laura~Ann Bauer, Anne Hultgren, Daniel~M. Silevitch, Li~Sun,
  et~al.
\newblock Magnetic alignment of fluorescent nanowires.
\newblock {\em Nano Letters\/}, 1:155--158, 2001.

\bibitem{Demirors2010electricdumbbells}
Ahmet~Faik Demir\"{o}rs, Patrick~M. Johnson, Carlos~M. van Kats, Alfons van
  Blaaderen, and Arnout Imhof.
\newblock Directed self-assembly of colloidal dumbbells with an electric field.
\newblock {\em Langmuir\/}, 26(18):14466--14471, 2010.

\bibitem{nelson2009helicalmagneticswimmer}
Li~Zhang, Jake~J. Abbott, Lixin Dong, Bradley~E. Kratochvil, Dominik Bell,
  et~al.
\newblock Artificial bacterial flagella: Fabrication and magnetic control.
\newblock {\em Applied Physics Letters\/}, 94(6):064107, 2009.

\bibitem{nelson2013MagneticSwimmerReview}
Kathrin~E. Peyer, Li~Zhang, and Bradley~J. Nelson.
\newblock Bio-inspired magnetic swimming microrobots for biomedical
  applications.
\newblock {\em Nanoscale\/}, 5:1259--1272, 2013.

\bibitem{KuzykYurke2008dielectrophoresis}
Anton Kuzyk, Bernard Yurke, J.~Jussi Toppari, Veikko Linko, and P{\"a}ivi
  T{\"o}rm{\"a}.
\newblock {D}ielectrophoretic trapping of {DNA} origami.
\newblock {\em Small\/}, 4(4):447--450, 2008.

\bibitem{toppari2015dielectrophoresisOrigami}
Boxuan Shen, Veikko Linko, Hendrik Dietz, and J.~Jussi Toppari.
\newblock Dielectrophoretic trapping of multilayer {DNA} origami nanostructures
  and {DNA} origami-induced local destruction of silicon dioxide.
\newblock {\em Electrophoresis\/}, 36(2):255--262, 2015.

\bibitem{mertig2016dielectrophoresisOrigami}
Anja Henning-Knechtel, Matthew Wiens, Mathias Lakatos, Andreas Heerwig, Frieder
  Ostermaier, et~al.
\newblock Dielectrophoresis of gold nanoparticles conjugated to {DNA} origami
  structures.
\newblock {\em Beilstein Journal of Nanotechnology\/}, 7:948--956, 2016.

\bibitem{DavisHowe2015antibodieselectricfield}
Sam Emaminejad, Mehdi Javanmard, Chaitanya Gupta, Shuai Chang, Ronald~W. Davis,
  et~al.
\newblock Tunable control of antibody immobilization using electric field.
\newblock {\em Proceedings of the National Academy of Sciences\/},
  112(7):1995--1999, 2015.

\bibitem{bensimon1995dnacombing}
D.~Bensimon, A.~J. Simon, V.~Croquette, and A.~Bensimon.
\newblock Stretching {DNA} with a receding meniscus: Experiments and models.
\newblock {\em Phys. Rev. Lett.\/}, 74:4754--4757, 1995.

\bibitem{schwartz1998dnaflowaligned}
Junping Jing, Jason Reed, John Huang, Xinghua Hu, Virginia Clarke, et~al.
\newblock Automated high resolution optical mapping using arrayed, fluid-fixed
  {DNA} molecules.
\newblock {\em Proc. Natl. Acad. Sci. USA\/}, 95:8046--8051, 1998.

\bibitem{li2013evaporativerediscovery}
Bo~Li, Wei Han, Myunghwan Byun, Lei Zhu, Qingze Zou, et~al.
\newblock Macroscopic highly aligned {DNA} nanowires created by controlled
  evaporative self-assembly.
\newblock {\em ACS Nano\/}, 7:4326--4333, 2013.

\bibitem{gothelf2015tubecombingalignment}
Bezu Teschome, Stefan Facsko, Kurt~V. Gothelf, and Adrian Keller.
\newblock Alignment of gold nanoparticle-decorated {DNA} origami nanotubes:
  Substrate prepatterning versus molecular combing.
\newblock {\em Langmuir\/}, 31(46):12823--12829, 2015.

\bibitem{dai2005germaniumwiresFluidAligned}
Dunwei Wang, Ryan Tu, Li~Zhang, and Hongjie Dai.
\newblock Deterministic one-to-one synthesis of germanium nanowires and
  individual gold nanoseed patterning for aligned nanowire arrays.
\newblock {\em Angewandte Chemie International Edition\/}, 44(19):2925--2929,
  2005.

\bibitem{yang2007alignedSilverWires}
Jiaxing Huang, Rong Fan, Stephen Connor, and Peidong Yang.
\newblock One-step patterning of aligned nanowire arrays by programmed dip
  coating.
\newblock {\em Angewandte Chemie International Edition\/}, 46(14):2414--2417,
  2007.

\bibitem{lieber2001orientednanowiresInP}
Yu~Huang, Xiangfeng Duan, Qingqiao Wei, and Charles~M. Lieber.
\newblock Directed assembly of one-dimensional nanostructures into functional
  networks.
\newblock {\em Science\/}, 291(5504):630--633, 2001.

\bibitem{li2015dnananowiresflow}
Bo~Li, Chuchu Zhang, Beibei Jiang, Wei Han, and Zhiqun Lin.
\newblock Flow-enabled self-assembly of large-scale aligned nanowires.
\newblock {\em Angewandte Chemie International Edition\/}, 54(14):4250--4254,
  2015.

\bibitem{guan2005pdmsstampcombing}
Jingjiao Guan and L.~James Lee.
\newblock Generating highly ordered {DNA} nanostrand arrays.
\newblock {\em Proc. Natl. Acad. Sci. U. S. A.\/}, 102:18321--18325, 2005.

\bibitem{horn1987absoluteorientation}
Berthold~K.P. Horn.
\newblock Closed-form solution of absolute orientation using unit quaternions.
\newblock {\em J. Opt. Sci. Am. A\/}, 4:629--642, 1987.

\bibitem{DietzCurvedOrigami2009}
H.~Dietz, S.M. Douglas, and W.M. Shih.
\newblock Folding {DNA} into twisted and curved nanoscale shapes.
\newblock {\em Science\/}, 325:725--730, 2009.

\bibitem{wei2013curvedSST}
Bryan Wei, Mingjie Dai, Cameron Myhrvold, Yonggang Ke, Ralf Jungmann, et~al.
\newblock Design space for complex {DNA} structures.
\newblock {\em Journal of the American Chemical Society\/},
  135(48):18080--18088, 2013.

\bibitem{marchi2014toward}
Alexandria~N. Marchi, Ishtiaq Saaem, Briana~N. Vogen, Stanley Brown, and
  Thomas~H. LaBean.
\newblock Toward larger {DNA} origami.
\newblock {\em Nano letters\/}, 14(10):5740--5747, 2014.

\bibitem{husale2010ssdna}
By~Sudhir Husale, Sangeeta Sahoo, Aleksandra Radenovic, Floriano Traversi,
  Paolo Annibale, et~al.
\newblock {ssDNA} binding reveals the atomic structure of graphene.
\newblock {\em Langmuir\/}, 26(23):18078--18082, 2010.

\bibitem{bohringer2004offsetringdesign}
Xiaorong Xiong, Sheng-Hsiung Liang, and K.F. B\"{o}hringer.
\newblock Geometric binding site design for surface-tension driven
  self-assembly.
\newblock In {\em 2004 IEEE International Conference on Robotics and
  Automation, 2004. Proceedings. {ICRA '04}.\/}, volume~2, pages 1141--1148.
  2004.

\bibitem{gopinath2016alignment}
Ashwin Gopinath, David Kirkpatrick, Paul Rothemund, and Chris Thachuk.
\newblock Progressive alignment of shapes.
\newblock In {\em Proceedings of the 28th Canadian Conference on Computational
  Geometry\/}, pages 230--236. 2016.

\bibitem{bohringer2004capillaryshapematching}
Sheng-Hsiung Liang, Xiaorong Xiong, and Karl~F. B\"{o}hringer.
\newblock Towards optimal designs for self-alignment in surface tension driven
  micro-assembly.
\newblock In {\em 17th IEEE International Conference on Micro Electro
  Mechanical Systems,{(MEMS 2004)}.\/}, pages 9--12. 2004.

\bibitem{spielmann1995TOTO1nmr90degrees}
H.~Peter Spielmann, David~E. Wemmer, and Jens~Peter Jacobsen.
\newblock Solution structure of a {DNA} complex with the fluorescent
  bis-intercalator {TOTO} determined by {NMR} spectroscopy.
\newblock {\em Biochemistry\/}, 34(27):8542--8553, 1995.

\bibitem{schins1999toto1at61egrees}
Juleon~M. Schins, Alexandra Agronskaia, Bart~G. de~Grooth, and Jan Greve.
\newblock Orientation of the chromophore dipoles in the {TOTO-DNA} system.
\newblock {\em Cytometry\/}, 37:230--237, 1999.

\bibitem{bennink1999yoyo1anisotropy69degrees}
Martin~L. Bennink, Orlando~D. Sch{\"a}rer, Roland Kanaar, Kumiko Sakata-Sogawa,
  Juleon~M. Schins, et~al.
\newblock Single-molecule manipulation of double-stranded {DNA} using optical
  tweezers: interaction studies of {DNA} with {RecA} and {YOYO-1}.
\newblock {\em Cytometry\/}, 36(3):200--208, 1999.

\bibitem{persson2009yoyo1anisotropy86degrees}
Fredrik Persson, Fredrik Westerlund, Jonas~O. Tegenfeldt, and Anders
  Kristensen.
\newblock Local conformation of confined {DNA} studied using emission
  polarization anisotropy.
\newblock {\em Small\/}, 5(2):190--193, 2009.

\bibitem{ha1999polarization}
Taekjip Ha, Ted~A. Laurence, Daniel~S. Chemla, and Shimon Weiss.
\newblock Polarization spectroscopy of single fluorescent molecules.
\newblock {\em The Journal of Physical Chemistry B\/}, 103(33):6839--6850,
  1999.

\bibitem{novotny2000singlemoleculepolarization}
B.~Sick, B.~Hecht, and L.~Novotny.
\newblock Orientational imaging of single molecules by annular illumination.
\newblock {\em Phys. Rev. Lett.\/}, 85:4482--4485, 2000.

\bibitem{BackerMoerner2016singlemoleculepolarization}
Adam~S. Backer, Maurice~Y. Lee, and W.~E. Moerner.
\newblock Enhanced {DNA} imaging using super-resolution microscopy and
  simultaneous single-molecule orientation measurements.
\newblock {\em Optica\/}, 3(6):659--666, 2016.

\bibitem{brasselet2016singlemoleculepolarization}
Cesar~Augusto Valades~Cruz, Haitham~Ahmed Shaban, Alla Kress, Nicolas Bertaux,
  Serge Monneret, et~al.
\newblock Quantitative nanoscale imaging of orientational order in biological
  filaments by polarized superresolution microscopy.
\newblock {\em Proceedings of the National Academy of Sciences\/},
  113(7):E820--E828, 2016.

\bibitem{milanovich1996twobiningmodesTOTO3}
N.~Milanovich, M.~Suh, R.~Jankowiak, G.~J. Small, and J.~M. Hayes.
\newblock Binding of {TO-PRO-3} and {TOTO-3} to {DNA}:  fluorescence and
  hole-burning studies.
\newblock {\em The Journal of Physical Chemistry\/}, 100(21):9181--9186, 1996.

\bibitem{pal2011rodsonorigami}
Suchetan Pal, Zhengtao Deng, Haining Wang, Shengli Zou, Yan Liu, et~al.
\newblock {DNA-}directed self-assembly of anisotropic plasmonic nanostructures.
\newblock {\em Journal of the American Chemical Society\/},
  133(44):17606--17609, 2011.

\bibitem{kuzyk2014rodsalign}
A.~Kuzyk, R.~Schreiber, H.~Zhang, A.~O. Govorov, T.~Liedl, et~al.
\newblock Reconfigurable {3D} plasmonic metamolecules.
\newblock {\em Nature Materials\/}, 13:862--866, 2014.

\bibitem{vybornyi2014porphyrinDNAcovalent}
Mykhailo Vybornyi, Alina~L. Nussbaumer, Simon~M. Langenegger, and Robert
  H{\"a}ner.
\newblock Assembling multiporphyrin stacks inside the {DNA} double helix.
\newblock {\em Bioconjugate Chemistry\/}, 25(10):1785--1793, 2014.

\bibitem{brongersma2012polarimeter}
Farzaneh Afshinmanesh, Justin~S. White, Wenshan Cai, and Mark~L. Brongersma.
\newblock Measurement of the polarization state of light using an integrated
  plasmonic polarimeter.
\newblock {\em Nanophotonics\/}, 1:125--129, 2012.

\bibitem{capasso2016metasurfacepolarimeter}
J.~P.~Balthasar Mueller, Kristjan Leosson, and Federico Capasso.
\newblock Ultracompact metasurface in-line polarimeter.
\newblock {\em Optica\/}, 3(1):42--47, 2016.

\bibitem{oldenbourg2016pnaspolarizationmicroscopy}
Shalin~B. Mehta, Molly McQuilken, Patrick~J. La~Riviere, Patricia Occhipinti,
  Amitabh Verma, et~al.
\newblock Dissection of molecular assembly dynamics by tracking orientation and
  position of single molecules in live cells.
\newblock {\em Proceedings of the National Academy of Sciences\/},
  113(42):E6352--E6361, 2016.

\bibitem{capasso2014metasurfaces}
Nanfang Yu and Federico Capasso.
\newblock Flat optics with designer metasurfaces.
\newblock {\em Nature Materials\/}, 13:139--150, 2014.

\bibitem{benson2011assembly}
O.~Benson.
\newblock Assembly of hybrid photonic architectures from nanophotonic
  constituents.
\newblock {\em Nature\/}, 480(7376):193--199, 2011.

\bibitem{ArmaniFraserFlaganVahala2007LabelFreeDetectionInMicrocavity}
A.~M. Armani, R.~P. Kulkarni, S.~E. Fraser, R.~C. Flagan, and K.~J. Vahala.
\newblock Label-free, single-molecule detection with optical microcavities.
\newblock {\em Science\/}, 317(5839):783--787, 2007.

\bibitem{hennessy2007stochasticQD}
K.~Hennessy, A.~Badolato, M.~Winger, D.~Gerace, M.~Atat{\"u}re, et~al.
\newblock Quantum nature of a strongly coupled single quantum dot--cavity
  system.
\newblock {\em Nature\/}, 445(7130):896--899, 2007.

\bibitem{riedrich2014deterministic}
Janine Riedrich-M{\"o}ller, Carsten Arend, Christoph Pauly, Frank M{\"u}cklich,
  Martin Fischer, et~al.
\newblock Deterministic coupling of a single silicon-vacancy color center to a
  photonic crystal cavity in diamond.
\newblock {\em Nano letters\/}, 14(9):5281--5287, 2014.

\bibitem{maitre2014measureandselectorientation}
Clotilde Lethiec, Julien Laverdant, Henri Vallon, Cl{\'e}mentine Javaux,
  Beno{\^i}t Dubertret, et~al.
\newblock Measurement of three-dimensional dipole orientation of a single
  fluorescent nanoemitter by emission polarization analysis.
\newblock {\em Phys. Rev. X\/}, 4:021037, 2014.

\bibitem{tartakovskii2013singlephotonstochasticQDonPCC}
I.J. Luxmoore, R.~Toro, O.~Del Pozo-Zamudio, N.A. Wasley, E.A. Chekhovich,
  et~al.
\newblock { III}--{V} quantum light source and cavity-{QED} on silicon.
\newblock {\em Scientific Reports\/}, 3:1239, 2013.

\bibitem{sapienza2015singlephotonstochasticQDonbragg}
L.~Sapienza, M.~Davanco, A.~Badolato, and K.~Srinivasan.
\newblock Nanoscale optical positioning of single quantum dots for bright and
  pure single-photon emission.
\newblock {\em Nature Communications\/}, 6:7833, 2015.

\bibitem{LyasotaKapon2015lithographicallydirectedQDs}
A.~Lyasota, S.~Borghardt, C.~Jarlov, B.~Dwir, P.~Gallo, et~al.
\newblock Integration of multiple site-controlled pyramidal quantum dot systems
  with photonic-crystal membrane cavities.
\newblock {\em Journal of Crystal Growth\/}, 414:192--195, 2015.

\bibitem{barth2010nanoassembled}
M.~Barth, S.~Schietinger, S.~Fischer, J.~Becker, N.~N{\"u}sse, et~al.
\newblock Nanoassembled plasmonic-photonic hybrid cavity for tailored
  light-matter coupling.
\newblock {\em Nano letters\/}, 10(3):891--895, 2010.

\bibitem{englund2010deterministic1DdiamondNV}
D.~Englund, B.~Shields, K.~Rivoire, F.~Hatami, J.~Vu{\v{c}}kovi{\'c}, et~al.
\newblock Deterministic coupling of a single nitrogen vacancy center to a
  photonic crystal cavity.
\newblock {\em Nano Letters\/}, 10(10):3922--3926, 2010.

\bibitem{sandoghdar2004terrylene}
R.J. Pfab, J.~Zimmermann, C.~Hettich, I.~Gerhardt, A.~Renn, et~al.
\newblock Aligned terrylene molecules in a spin-coated ultrathin crystalline
  film of p-terphenyl.
\newblock {\em Chemical Physics Letters\/}, 387:490--495, 2004.

\bibitem{ToninelliSandoghdar2010alignedmoleculesNearIR}
C.~Toninelli, K.~Early, J.~Bremi, A.~Renn, S.~G\"{o}tzinger, et~al.
\newblock Near-infrared single-photons from aligned molecules in ultrathin
  crystalline films at room temperature.
\newblock {\em Opt. Express\/}, 18(7):6577--6582, 2010.

\bibitem{Polisseni2016aligneddibenzoterryleneemitters}
Claudio Polisseni, Kyle~D. Major, Sebastien Boissier, Samuele Grandi, Alex~S.
  Clark, et~al.
\newblock Stable, single-photon emitter in a thin organic crystal for
  application to quantum-photonic devices.
\newblock {\em Opt. Express\/}, 24(5):5615--5627, 2016.

\bibitem{LesikJacques2014alignedNVcenters}
M.~Lesik, J.-P. Tetienne, A.~Tallaire, J.~Achard, V.~Mille, et~al.
\newblock Perfect preferential orientation of nitrogen-vacancy defects in a
  synthetic diamond sample.
\newblock {\em Applied Physics Letters\/}, 104(11):113107, 2014.

\bibitem{MichlWrachtrup2014alignedNVcenters}
Julia Michl, Tokuyuki Teraji, Sebastian Zaiser, Ingmar Jakobi, Gerald Waldherr,
  et~al.
\newblock Perfect alignment and preferential orientation of nitrogen-vacancy
  centers during chemical vapor deposition diamond growth on (111) surfaces.
\newblock {\em Applied Physics Letters\/}, 104(10):102407, 2014.

\bibitem{LimBardeen2004TetraceneSuperradianceInThinFilms}
Sang-Hyun Lim, Thomas~G. Bjorklund, Frank~C. Spano, and Christopher~J. Bardeen.
\newblock Exciton delocalization and superradiance in tetracene thin films and
  nanoaggregates.
\newblock {\em Phys. Rev. Lett.\/}, 92:107402, Mar 2004.

\bibitem{HettichSandoghdar2002coupledalignedMolecules}
C.~Hettich, C.~Schmitt, J.~Zitzmann, S.~K{\"u}hn, I.~Gerhardt, et~al.
\newblock Nanometer resolution and coherent optical dipole coupling of two
  individual molecules.
\newblock {\em Science\/}, 298(5592):385--389, 2002.

\bibitem{Novotny2011PhysicsTodayNearFieldOptics}
Lukas Novotny.
\newblock From near-field optics to optical antennas.
\newblock {\em Physics Today\/}, 64(7):47--52, 2011.

\bibitem{SalandrinoEngheta2007OpticalNanocircuitsPhysical}
Andrea Al\`{u}, Alessandro Salandrino, and Nader Engheta.
\newblock Parallel, series, and intermediate interconnections of optical
  nanocircuit elements. 2. nanocircuit and physical interpretation.
\newblock {\em J. Opt. Soc. Am. B\/}, 24(12):3014--3022, 2007.

\bibitem{AluEngheta2008OpticalNanoantennasNanocircuits}
Andrea Al\`{u} and Nader Engheta.
\newblock Tuning the scattering response of optical nanoantennas with
  nanocircuit loads.
\newblock {\em Nature Photonics\/}, 2:307--310, 2008.

\bibitem{woo2014cation}
Sungwook Woo and Paul~W.K. Rothemund.
\newblock Self-assembly of two-dimensional {DNA} origami lattices using
  cation-controlled surface diffusion.
\newblock {\em Nature Communications\/}, 5:4889, 2014.

\bibitem{decker2007ChiralMetasurfaces}
M.~Decker, M.~W. Klein, M.~Wegener, and S.~Linden.
\newblock Circular dichroism of planar chiral magnetic metamaterials.
\newblock {\em Opt. Lett.\/}, 32(7):856--858, 2007.

\bibitem{ye2016MonlinearMetasurfaceHolographyCD}
Weimin Ye, Franziska Zeuner, Xin Li, Bernhard Reineke, Shan He, et~al.
\newblock Spin and wavelength multiplexed nonlinear metasurface holography.
\newblock {\em Nature Communications\/}, 7:11930, 2016.

\bibitem{ratner1974rectifier}
Arieh Aviram and Mark~A. Ratner.
\newblock Molecular rectifiers.
\newblock {\em Chemical Physics Letters\/}, 29:277--283, 1974.

\bibitem{ratner2015newrectifier}
Colin~Van Dyck and Mark~A. Ratner.
\newblock Molecular rectifiers: A new design based on asymmetric anchoring
  moieties.
\newblock {\em Nano Letters\/}, 15(3):1577--1584, 2015.

\bibitem{wong2013CNTcomputer}
Max~M. Shulaker, Gage Hills, Nishant Patil, Hai Wei, Hong-Yu Chen, et~al.
\newblock Carbon nanotube computer.
\newblock {\em Nature\/}, 501:526--530, 2013.

\bibitem{origamiPEGpurification2014}
E.~Stahl, T.G. Martin, F.~Praetorius, and H.~Dietz.
\newblock {F}acile and scalable preparation of pure and dense {DNA} origami
  solutions.
\newblock {\em Angew. Chem. Int. Edit.\/}, 53(47):12735--12740, 2014.

\bibitem{hogberg2015OrigamiPurification}
Alan Shaw, Erik Benson, and Bj\"{o}rn H\"{o}gberg.
\newblock Purification of functionalized {DNA} origami nanostructures.
\newblock {\em ACS Nano\/}, 9(5):4968--4975, 2015.

\bibitem{noy2007handbook}
A.~Noy.
\newblock {\em Handbook of Molecular Force Spectroscopy\/}.
\newblock Springer, New York, 2007.

\bibitem{hogberg2012intercalatordistortion}
Yong-Xing Zhao, Alan Shaw, Xianghui Zeng, Erik Benson, Andreas~M. Nystr\"{o}m,
  et~al.
\newblock {DNA} origami delivery system for cancer therapy with tunable release
  properties.
\newblock {\em ACS Nano\/}, 6(10):8684--8691, 2012.

\bibitem{choi2017intercalatororigamiswitching}
Haorong Chen, Ruixin Li, Shiming Li, Joakim Andr\'{e}asson, and Jong~Hyun Choi.
\newblock Conformational effects of {UV} light on {DNA} origami.
\newblock {\em Journal of the American Chemical Society\/}, 139(4):1380--1383,
  2017.
\newblock PMID: 28094518.

\bibitem{ke2012intercalatingunderwinding}
Yonggang Ke, Ga\"{e}tan Bellot, Niels~V. Voigt, Elena Fradkov, and William~M.
  Shih.
\newblock Two design strategies for enhancement of multilayer--{DNA}--origami
  folding: underwinding for specific intercalator rescue and staple-break
  positioning.
\newblock {\em Chemical science\/}, 3:2587--2597, 2012.

\bibitem{Wang2009niceNanowireDSAreview}
Michael~C.P. Wang and Byron~D. Gates.
\newblock Directed assembly of nanowires.
\newblock {\em Materials Today\/}, 12(5):34--43, 2009.

\bibitem{arbona2012interhelixGapCrossovers}
Jean~Michel Arbona, Jean-Pierre Aim\'{e}, and Juan Elezgaray.
\newblock Modeling the mechanical properties of {DNA} nanostructures.
\newblock {\em Phys. Rev. E\/}, 86:051912, Nov 2012.

\end{thebibliography}
\bibliographystyle{myunsrt}

\end{document}


\maketitle

\begin{center}
$^{*}$Corresponding author. Email: ashwing@caltech.edu (A.G.); pwkr@dna.caltech.edu (P.W.K.R)
\end{center}

\vspace*{0.5in}
{\bf This PDF file includes:}
\vspace*{0.2in}

\begin{figure}[htp]
\includegraphics[width=\textwidth]{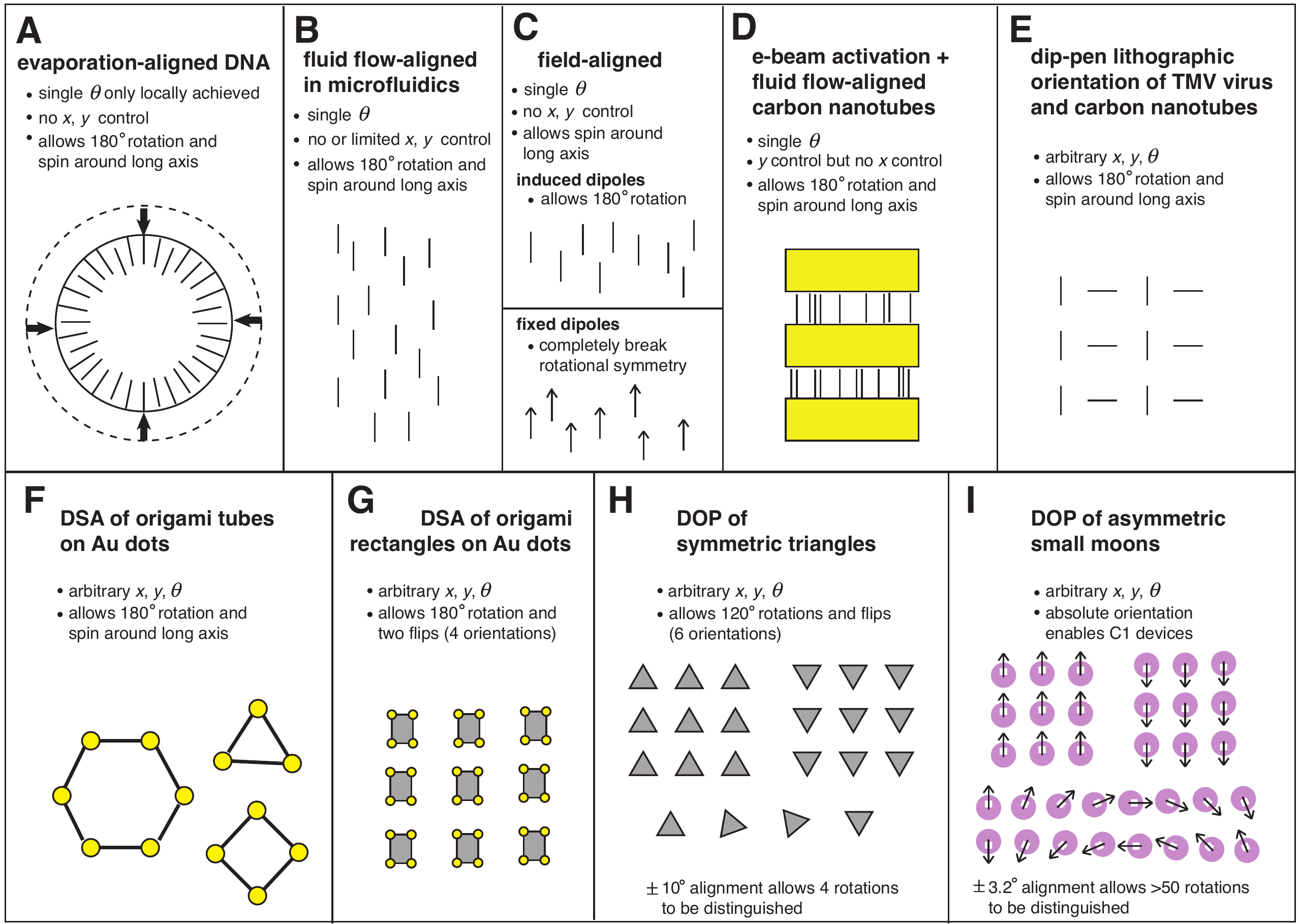}
\caption{{\bf Comparison of methods for aligning nanodevices.} Here we
  give representative schematics for a wide variety of techniques
  which could be used to align nanodevices comprised of, or templated
  on, DNA strands, carbon nanotubes, metal nanowires, and DNA
  origami. We do not review a wide body of work which deals with
  positioning spherical particles
  (e.g. [{\em 11, 12, 15}] )
 since we are interested in
  work which has the ability to perform orientation of at least
  dipoles, if not absolute orientation. We note that in general, one
  dimensional structures cannot be used for absolute orientation
  because they can spin arbitrarily along their long axis. ({\bf A})
  Simple flow powered by the receding meniscus of an evaporating drop ({\em 38-41}),
 often termed molecular combing,
  has long been used to arrange DNA and other one-dimensional DNA
  nanostructures, aligning them to a single $\theta$, at least
  locally. Inorganic nanowires have been aligned as well ({\em 42, 43}).
({\bf B}) Combined with microfluidics ({\em 44}),
 shear from moving experimental setups ({\em 45}),
and a variety of
  stamping and pattern-transfer methods, flow alignment can be made
  considerably more powerful, and allowing limited control over the
  $x$-$y$ positioning of one dimensional nanostructures. While a
  single application is still limited to a single $\theta$, multiple
  applications can lead to arrays of crossed structures ({\em 44, 46}).
  Similar results for other techniques
  such as Langmuir Blodgett films are reviewed in ({\em 103}).
({\bf C}) Magnetic and
  electric fields have been used to align carbon nanotubes ({\em 27, 28})
  and metallic nanowires ({\em 29, 30}),
and particulate dumbbells ({\em 31}).
In these examples, alignment
  forces act on induced dipoles, devices so aligned are subject to
  random 180\degree\ rotation. Alignment of fixed dipoles, for example
  the electrostatic dipoles of antibodies ({\em 37}),
or the magnetic dipoles
  of microfabricated helical swimmers ({\em 32, 33})
  allow rotational symmetry to be broken.  We neglect to draw schema
  field based dielectrophoretic methods which could potentially
  achieve arbitrary $x$,$y$ control with intricate electrode patterns ({\em 34--36});
  however, orientation at small electrode gaps tends to be poorer than
  for large-scale uniform fields.  ({\bf D}) A combination of
  chemical differentiation (via e-beam activation) and flow alignment can achieve
  orientation (up to 180\degree\ rotation) and some control over
  position  ({\em 18}).
 ({\bf E}) Scanning probe-based
  chemical differentiation of a surface (here dip-pen nanolithography)
  allows linear viruses ({\em 16})
or carbon  nanotubes  ({\em 17})
to be oriented
  arbitrarily.  ({\bf F}) Lithographic patterning of gold dots allows
  linear DNA structures terminated with thiols to be arbitrarily
  oriented ({\em 19, 21})
  similar work on block copolymers ({\em 20})
compromises arbitrary  $x$, $y$, $\theta$ control for potential scalability.  ({\bf G})
  Extension of the gold-dot/thiol approach to 2D nanostructures
  (rectangles) allows orientational freedom to be limited to just four
  degenerate orientations ({\em 22}).
({\bf H}) DNA  origami placement of equilateral triangles still leaves six
  degenerate orientations, and orientational fidelity is relatively
  coarse, allowing only four rotations to be distinguished ({\em 24--26}).
({\bf I}) The current work with
  asymmetric small moons achieves absolute and arbitrary orientation,
  and should enable more than 50 distinguishable rotations.}
\label{fig:AlignmentComparison}
\end{figure}

\noindent  \hspace*{0.5in} Materials and Methods \\
\hspace*{0.5in} Figs. S\ref{fig:AlignmentComparison} through S\ref{fig:PCCSEM} \\

\newpage

\noindent {\bf \LARGE Contents}  
\bvsp

\noindent {\bf Materials and Methods} \dotfill \pageref{MM} \\
\indent DNA origami designs, preparation and purification \dotfill \pageref{origami-design}\\
\indent Placement chip fabrication \dotfill \pageref{chip-fabrication}\\
\indent Photonic crystal fabrication \dotfill \pageref{pcc-fabrication}\\
\indent FDTD simulations of PCCs \dotfill \pageref{FDTD}\\
\indent Origami placement experiments \dotfill \pageref{placement-experiments}\\
\indent Ethanol drying \dotfill \pageref{ethanol-drying}\\
\indent Troubleshooting placement experiments \dotfill \pageref{troubleshooting}\\
\indent AFM characterization \dotfill \pageref{afm-characterization}\\
\indent TOTO-3 binding and optical experiments \dotfill \pageref{afm-characterization}\\
\noindent {\bf Figs. S\ref{fig:AlignmentComparison} through S\ref{fig:PCCSEM}} \dotfill \pageref{fig:RRTPlaced} \\

\setcounter{secnumdepth}{-2}  

\newpage

\label{MM}
\section{Materials and Methods}

\label{origami-design}
\subsection{DNA origami designs, preparation and purification}
\noindent {\em Designs.} Here, all origami were designed with caDNAno
({\small \tt http://cadnano.org/}) to position all staple ends on the
same face of the origami so that single-stranded 20T extensions to 5'
staple ends would all project from the same face of the origami.  All
caDNAno design files and lists of staples are included as a
supplementary zip archive: \shsp {\tt \small AA-designs+scripts.zip}.
For right triangle designs, we list two versions of each staple: one is
as designed from caDNAno and the other is with 20T extension on the 5'
end. The three origami used in this work are as follows:

\begin{enumerate}

\item {\em Right-handed right triangle (RRT):} Staples on the right-hand face of this triangle 
   were extended. The caDNAno design and staple list files
   are {\tt \small RRT.json}, {\tt \small RRT-Staples.xls} and {\tt \small RRT-T20-Staples.xls}.

\item {\em Left-handed right triangle (LRT):} This design is similar
  to that for the right-handed right triangle, except that staple ends
  have been shifted by half a DNA turn so that they fall onto the
  left-hand face of the triangle.  The caDNAno design and staple list
  files are {\tt \small LRT.json}, {\tt \small LRT-Staples.xls} and
  {\tt \small LRT-T20-Staples.xls}.

\item {\em Small moon:} CaDNAno design and staple list files are {\tt
  \small small-moon.json} and {\tt \small small-moon-staples.xslx}; staples
  are extended with 20T on their 5' ends.

\end{enumerate}

\noindent {\em Preparation.} Staple strands (Integrated DNA Technologies,
100~$\mu$M each in water) and the scaffold strand (single-stranded
M13mp18, 400~nM from Bayou Biolabs for right triangles; p8064, 100~nM
from Tilibit for small moons) were mixed together to target
concentrations of 100~nM (each staple) and 40~nM, respectively (a
2.5:1 staple:scaffold ratio) in 10~mM Tris Base, 1~mM EDTA buffer
(adjusted to pH~8.35 with HCl) with 12.5~mM magnesium chloride
(TE/Mg\textsuperscript{2+}). 50~$\mu$L volumes of staple/scaffold
mixture were heated to 90\degree C for 5~min and annealed from
90\degree C to 20\degree C at -0.2\degree C/min in a PCR machine. We
used 0.5~ml DNA LoBind tubes (Eppendorf) to minimize loss of origami
to the sides of the tube.

\vsp \Dangersign \shsp \parbox{6in}{Do not use acetate in preparation
  of the formation buffer for DNA origami ({\em e.g.} using acetic
  acid to adjust pH). For historical reasons acetate-containing
  TAE/Mg\textsuperscript{2+}, a gel electrophoresis buffer, has been
  used for preparing DNA origami. In the context of origami placement,
  acetate ions cause a high background of small particles to appear,
  presumably insoluble acetate salts.} 

\vsp \noindent Recipes of all origami used in this paper:
\begin{center}
\label{my-label}
\begin{tabular}{lllll}
          & Scaffold  & Staple            & 10x Buffer    & Water \\
RRT, 0\% T      & 5 \uL\ (M13mp18)     & 16 \uL\ (RRT)         & 5 \uL\        & 24 \uL\  \\ 
RRT, 12.5\% T   & 5 \uL\      & 2 \uL\  (RRT-20T) + 14 \uL\ (RRT) & 5 \uL\        & 24 \uL\  \\ 
RRT, 25\% T    & 5 \uL\      & 4 \uL\  (RRT-20T) + 12 \uL\ (RRT) & 5 \uL\        & 24 \uL\  \\ 
RRT, 37.5\% T   & 5 \uL\      & 6 \uL\  (RRT-20T) + 10 \uL\ (RRT) & 5 \uL\        & 24 \uL\  \\ 
RRT, 50\% T    & 5 \uL\      & 8 \uL\  (RRT-20T) + 8 \uL\ (RRT)  & 5 \uL\        & 24 \uL\  \\ 
RRT, 62.5\% T   & 5 \uL\      & 10 \uL\  (RRT-20T) + 6 \uL\ (RRT) & 5 \uL\        & 24 \uL\  \\ 
RRT, 75\% T     & 5 \uL\      & 12 \uL\  (RRT-20T) + 4 \uL\ (RRT) & 5 \uL\        & 24 \uL\  \\ 
RRT, 87.5\% T   & 5 \uL\      & 14 \uL\  (RRT-20T) + 2 \uL\ (RRT) & 5 \uL\        & 24 \uL\  \\ 
RRT, 100\% T    & 5 \uL\      & 16 \uL\  (RRT-20T)            & 5 \uL\        & 24 \uL\  \\ 
LRT, 0\% T      & 5 \uL\      & 16 \uL\ (LRT)         & 5 \uL\        & 24 \uL\  \\ 
LRT, 12.5\% T   & 5 \uL\      & 2 \uL\  (LRT-20T) + 14 \uL\ (LRT) & 5 \uL\        & 24 \uL\  \\ 
LRT, 25\% T     & 5 \uL\      & 4 \uL\  (LRT-20T) + 12 \uL\ (LRT) & 5 \uL\        & 24 \uL\  \\ 
LRT, 37.5\% T   & 5 \uL\      & 6 \uL\  (LRT-20T) + 10 \uL\ (LRT) & 5 \uL\        & 24 \uL\  \\ 
LRT, 50\% T     & 5 \uL\      & 8 \uL\  (LRT-20T) + 8 \uL\ (LRT)  & 5 \uL\        & 24 \uL\  \\ 
LRT, 62.5\% T   & 5 \uL\      & 10 \uL\  (LRT-20T) + 6 \uL\ (LRT) & 5 \uL\        & 24 \uL\  \\ 
LRT, 75\% T     & 5 \uL\      & 12 \uL\  (LRT-20T) + 4 \uL\ (LRT) & 5 \uL\        & 24 \uL\  \\ 
LRT, 87.5\% T   & 5 \uL\      & 14 \uL\  (LRT-20T) + 2 \uL\ (LRT) & 5 \uL\        & 24 \uL\  \\ 
LRT, 100\% T    & 5 \uL\      & 16 \uL\  (LRT-20T)        & 5 \uL\        & 24 \uL\  \\ 
Small moon & 20 \uL\  (p8064) & 10 \uL\ (20T modified)   & 5 \uL\        & 15 \uL\   
\end{tabular}
\end{center}


\newpage

\vsp \noindent {\em Purification.} A high concentration of excess
staples will prevent origami placement. Thus origami were purified
away from excess staples using 100~kD molecular weight cut-off filters
spin filters (Amicon Ultra-0.5 Centrifugal Filter Units with
Ultracel-100 membranes, Millipore, UFC510024). By the protocol below,
recovery is generally 40--50\% and staples are no longer visible by
agarose gel:

\begin{enumerate}
\item Wet the filter by adding 500~\uL\ TE/Mg\textsuperscript{2+}.
\item Spin filter at 2000~rcf for 6~min at 4\Tp, until the volume in
  the filter is 50~\uL. Discard the filtrate.
\item Add 50~\uL\ of unpurified origami and
  400~\uL\ TE/Mg\textsuperscript{2+}. Spin at 2000~rcf for 6~min at
  4\Tp.
\item Discard the filtrate. Add 450~\uL\ TE/Mg\textsuperscript{2+} and
  spin at 2000~rcf for 6~min at 4\Tp.
\item Repeat step (4) three more times.
\item Invert the filter onto a clean tube and spin at 2000~rcf for
  6~min at \Tp\ to collect purified origami ($\sim 50$\uL).
\end{enumerate}

\noindent Total time for this purification is roughly
40~minutes. Post-purification, origami are quantified using a NanoDrop
spectrophoto- meter (Thermo Scientific), estimating the molar extinction
coefficient of the DNA origami as that of a fully double-stranded
M13mp18 molecule ($\epsilon =$123,735,380/M/cm; we do not correct for
small single-stranded loops which are present on the edges of some
designs).  We typically work with stock solutions of 15--20~nM DNA
origami (2--2.5 OD). The working concentration for origami during
placement is 100~pM, which is too small to be measured with the
NanoDrop, so serial dilutions must be performed. High quality
placement is very sensitive to origami concentration. To maintain
consistency for each series of experiments for a particular shape, a
single high concentration stock solution (from a single purification)
was maintained and diluted to a nominal concentration of 100~pM as
needed.

\textcolor{Black}{\textbf{Note:} All of the work reported in this
  paper was performed with spin-column purified origami, which is
  suitable for small amounts of origami. Larger-scale purification can
  be achieved using PEG precipitation ({\em 97});
  we have performed placement
  experiments using PEG-purified origami, and achieved good results.
  See ref.~({\em 98}) 
  for other large-scale
  purification techniques and a comparison of their efficiency.}

\vsp
\Dangersign \shsp \parbox{6in}{After purification and
  quantification, it is especially important to use DNA LoBind tubes
  (Eppendorf) for storage and dilution of low concentration DNA
  origami solutions. Low dilutions, {\em e.g.} 100~pM, must be made
  fresh from more concentrated solutions and used immediately---even
  overnight storage can result in total loss of origami to the sides
  of the tube. Addition of significant amounts of carrier DNA to
  prevent origami loss may prevent origami placement, just as excess
  staples do. We have not yet determined whether other blocking agents
  such as BSA might both prevent origami loss and preserve placement.}

\newpage

\label{chip-fabrication}
\subsection{Fabrication of binding sites}
Fabrication of binding sites is very similar to that found in 
({\em 8}) 
and ({\em 26}) 
here we give an overview of the
process and a couple places where it departs from previous work. All
steps were carried out in Caltech's Kavli Nanoscience Institute
cleanroom.

For non-PCC exmperiments, fabrication begins with a thermally-grown
SiO\textsubscript{2} layer (on a silicon wafer) which is cleaned and
silanized with a trimethyl silyl passivation layer by vapor deposition
of HMDS (hexamethyldisilazane). A thin (80~nm) layer of PMMA~950~A2
(MicroChem Corp.; our previous work used a thicker layer of
PMMA~950~A3) is spun-coat on the substrate as a resist. Binding sites
in the shape of a DNA origami are defined in the resist with e-beam
lithography and developed. After the binding sites are defined, the
trimethyl silyl passivation layer is selectively removed at the
binding sites using an anisotropic O\textsubscript{2}-plasma etch, in
a process we term `activation'. Finally, the residual PMMA resist is
removed to reveal a substrate that is composed of two chemically
distinct regions: (i) origami-shaped features covered with ionizable
surface silanols (-OH) and (ii) a neutrally-charged background covered
with trimethyl silyl groups. This procedure enables good placement in
35~mM~Mg\textsuperscript{2+}.

For the photonic crystal experiments on silicon nitride, the complex
geometry of the holes and membranes means that we cannot add an HMDS
passivation layer to some surfaces. To avoid nonspecific binding of
origami to these surfaces, we perform DOP at a lower
Mg\textsuperscript{2+} concentration of 12.5~mM.  To achieve strong
adhesion to binding sites under this condition, we silanize activated
sites with 0.1\% CTES (carboxyethylsilanetriol from Gelest, 25\% w/v
Catalog \# SIC2263.0) in 10~mM Tris, pH~8.0 for 30 minutes {\em
  before} the resist is stripped. In our previous work ({\em 8}),
silanization was performed with lower
concentration CTES (0.01\% for 10 minutes) {\em after} the resist was
stripped but the new protocol results in lower background binding
since the HMDS passivation layer is protected beneath the resist
during silanization.

\label{pcc-fabrication}
\subsection{Fabrication of PCC arrays}
Here, fabrication of PCC arrays is very similar to the process found
in ({\em 8}) 
for ``isolated PCCs'', rather than
the process for ``close-packed arrays''; this is because the PCC arrays
described here are smaller and do not justify the more complex process
used to fabricate very large, suspended arrays of PCCs.  All steps
were carried out in Caltech's Kavli Nanoscience Institute cleanroom.

A schematic of the fabrication process is shown in Fig.~S13 and SEM of
the result in Fig.~S14. Fabrication began with double-side polished
silicon wafers (DSP, $\langle$100$\rangle$, 380$\pm$10$\mu$m thick,
University Wafers, Rogue Valley Microdevices) with 275~nm layers of
LPCVD-grown SiN on both sides of each wafer. The wafer was cleaned and
alignment markers were defined in the SiN layer by e-beam lithography
and modified-Bosch ICP etching.  The substrate was then cleaned and
silanized with a trimethyl silyl passivation layer using vapor
deposition of HMDS. Next, binding sites in the shape of a DNA origami
were defined using e-beam lithography at specific locations on the
front face using the previously-defined alignment markers. Binding
sites were then activated with a short O\textsubscript{2} plasma etch
to create silanols, the silanols were converted to carboxyl groups
(see ``Fabrication of binding sites''), and the resist was stripped.
New resist was spun on, and PCCs were defined around binding site by
e-beam lithography and modified-Bosch ICP etching of the SiN
layer. Finally, PCCs were suspended using a XeF\textsubscript{2}
isotropic etch of the underlying Si layer.

\label{FDTD}
\subsection{FDTD simulations of PCCs}
Three dimensional (3D) finite difference time domain (FDTD) simulation
was used both for PCC design and to generate simulated LDOS for
comparison with experimental maps of the resonant cavity modes.  All
simulations were performed using {\em FDTD Solutions} from Lumerical
Solutions, Inc {\tt \small https://www.lumerical.com/}. Lumerical simulation
files can be found in the directory {\tt \small LumericalScripts} in the zip archive
{\tt \small AA-designs+scripts.zip}. Matlab files for creating Autocad versions of optimized
resonators can be found in the directory {\tt \small AutocadScriptGenerator} in the
same zip archive.

To design the photonic crystal we fixed the refractive index of SiN at
2.05, the thickness of the SiN membrane at 275~nm, and adjusted $r$,
$r/a$, $r1$, $r2$ and $s$ (inset, Fig.~S14A) to maximize quality factor
within the wavelength range of 655--660~nm.  Photonic crystal size was
set to $20a$ in the $x$ direction and $34.64a$ in the $y$ direction.
Boundary conditions were implemented by introducing a perfect matching
layer around the structure.  The simulation discretization was set to
$a/R$ in the $x$-direction, $0.866a/R$ in the $y$-direction, and $a/R$
in the $z$-direction, where the variable $R$ was set to 10 for PCC
design (so that PCC parameter could be quickly optimized), and set to
20 to generated simulated LDOS of higher resolution for comparison
with experimental mode maps.  The simulation modeled emission from a
single dipole with polarization $P(x,y,z) = (1,1,0)$, located at a
weak symmetry point close the cavity surface.

\newpage 
\label{placement-experiments}
\subsection{Origami placement experiments}

Below we describe the placement protocol in four steps. See
troubleshooting guide on page~\pageref{troubleshooting} for an
enumeration of problems and suggestions.  See our previous work ({\em
  26})
for a greater discussion of
origami placement; the supplemental material for that work provides a
figure (Fig.~S3) showing how substrates should look during the placement
process. 

\begin{enumerate}
\item \textbf{Binding.} A 50~mm petri dish was prepared with a
  moistened kimwipe to limit evaporation. For non-PCC samples,
  solution with 100~pM origami was prepared in \textbf{placement
    buffer} (10~mM Tris, 35~mM Mg\textsuperscript{2+}, pH~8.3) and a
  20~\uL\ drop was deposited in the middle of the chip on top of the
  patterned region. For PCC arrays, 12.5~mM Mg\textsuperscript{2+} was
  used in the placement buffer (see note below). The chip was placed
  in a closed, humid petri dish and the origami solution was allowed
  to incubate on the chip for 1~hour.

\item \textbf{Initial wash.} After the 1~hour incubation, excess
  origami (in solution) were washed away with at least 8~buffer washes
  by pipetting 60~\uL\ of fresh \textbf{placement buffer} onto the
  chip, and pipetting 60~\uL\ off of the chip. Each of the
  8~washes consisted of pipetting the 60~\uL\ volume up and down 2--3
  times to \textbf{mix} the fresh buffer with existing buffer on the
  chip. This initial wash took about 2~minutes.

\item \textbf{Tween wash.} Next, in order to remove origami that were
  non-specifically bound to the passivated background, the chip was
  buffer-washed 5~times using a \textbf{Tween washing buffer} made by
  adding 0.1\% Tween 20 (v/v) to placement buffer. This took about
  1~minute. Because of the low surface tension of the Tween washing
  buffer, these washes were somewhat tricky: they involve adding
  20--40~\uL\ of tween wash buffer, just enough to cover most of the
  chip, but not enough to spill over the chip and wet the back side of
  the chip (this may introduce dust contamination from the petri
  dish). After the 5th wash, the chip was left to incubate for
  30~minutes.

\item \textbf{Final wash.} Lastly, the chip was buffer-washed 8~times
  back into either a higher pH \textbf{stabilizing buffer} for wet AFM
  imaging (10~mM Tris, 35~mM Mg\textsuperscript{2+}, pH~8.9; this
  prevents movement during AFM) or placement buffer for subsequent
  drying.  This took about 2~minutes. These washes were high volume
  (60~\uL) and were intended to completely remove the Tween~20. The
  amount of Tween~20 left was monitored qualitatively by the surface
  tension of the drop (roughly, by eye). When a 20\uL\ drop covered
  roughly the same area as the initially deposited drop, it was
  assumed that the Tween~20 had been sufficiently removed. After the
  last wash, the chip was left with roughly 20~\uL\ of buffer and was
  ready for AFM imaging or drying.

\end{enumerate}

\Dangersign \shsp \parbox{6in}{Do not use EDTA in placement, Tween washing,
  or imaging buffers. It is unnecessary in this context, and will
  slightly change the effective Mg\textsuperscript{2+} concentration
  available for placement.}
\vsp

\Dangersign \shsp \parbox{6in}{Do not allow the patterned region with
  binding sites to dry at any point during the binding step or
  subsequent buffer washes. Inadvertent dewetting of the binding sites
  leads to distortion of the origami (causing them to ball up) as well
  as the formation of salt crystals on the binding sites. If the
  substrate needs to be dried follow the ethanol drying procedure
  presented in the next section.}
\vsp

\Dangersign \shsp \parbox{6in}{Use Tween~20, rather than other
  surfactants. Tween~80 and SDS, which are two other common
  surfactants, lead to very different results. Tween~80 leads to the
  total removal of placed origami from the substrate. SDS does not
  remove excess origami from the trimethyl silyl background.}
\vsp

\Dangersign \shsp \parbox{6in}{Make sure that chips are not exposed to Tween~20 until
            {\em after} the origami have been deposited. Tween~20
            applied before binding significantly reduces binding to
            activated sites.}
\vsp

\Dangersign \shsp \parbox{6in}{Make fresh buffer solutions every
  week. Here and elsewhere in this work, we use buffers at low
  strength (typically 10~mM) to minimize background binding and to make
  complete washing into different buffers easier. This means the
  buffers have low buffering capacity and the pH will decrease with time (and
  placement may cease to work).}
\vsp

{\bf Note:} For non-PCC samples the binding of DNA origami to
SiO\textsubscript{2} is mediated by Mg\textsuperscript{2+} binding to
surface silanols.  For PCC samples, the origami binding is mediated by
Mg\textsuperscript{2+} binding to carboxyl groups generated by CTES
silanization.  The use of carboxylated binding sites allows
high-quality origami placement and orientation on SiN PCC membranes at
a much lower Mg\textsuperscript{2+} concentration (12.5~mM) than that
required (35~mM) for O\textsubscript{2} plasma-activated binding sites
on SiO\textsubscript{2}. We suggest that the effect is due to the
difference in pK\textsubscript{a} between these two functional groups:
similar surface carboxyl groups ({\em 99})
have a pK\textsubscript{a}$\sim$6, while silanol groups have a
pK\textsubscript{a} of 8.3. Thus binding sites with carboxyl groups
should carry a higher negative charge at our working pH of 8.3, they
should bind more Mg\textsuperscript{2+}, and should enable the
observed binding of origami at lower Mg\textsuperscript{2+}
concentration. In addition to decreasing the potential for salt
artifacts during drying, the use of carboxyl groups has a further very
important added benefit. During the extensive PCC fabrication process,
different surface types as identified by a specific series of
treatments, are created. Some of these, for example the inside of the
PCC holes or the back side of the PCC membranes, are not passivated
with trimethyl silyl groups, and appear to bind some origami at higher
Mg\textsuperscript{2+} concentrations.  Thus the use of carboxylated
binding sites (and hence a lower Mg\textsuperscript{2+} concentraton
for placement) decreases nonspecific origami binding and ensures that
under our buffer conditions the only locations at which origami can
stably bind are the intended binding sites.

\label{ethanol-drying}
\subsection{Ethanol drying}
After DNA origami were immobilized on chips (and potentially labeled
with TOTO-3), they were dried by exposure to an ethanol dilution
series: 10 seconds in 50\% ethanol, 30 seconds in 75\% ethanol, and
120 seconds in 90\% ethanol. To remove remaining 90\% ethanol, chips
were air dried.

\vsp
\Dangersign \shsp \parbox{6in}{If arrays of placed origami are subjected to solutions
with less than 80\% ethanol for an extended period ($> 2$~minutes), a
significant reduction in binding is observed.}

\vsp
\Dangersign \shsp \parbox{6in}{Drying with stream of N\textsubscript{2} can lead to drying
artifacts (e.g. micron-scale streaks visible via AFM).}

\newpage

\label{troubleshooting}
\subsection{Troubleshooting origami placement}

\begin{center}
    \begin{tabular}{ | m{4cm} | p{4.5cm} | p{6.5cm} |}
    \hline
    Problem & Likely cause & Solution \\ \hline

\vsp \mbox{Site occupancy below 90\%.} & 
  \mbox{$\bullet$ Old chip with inactive sites.} \newline
  \mbox{$\bullet$ Low origami concentration.}   \newline
    \newline
  \mbox{$\bullet$ Short incubation time.} \newline
  \mbox{$\bullet$ Low Mg\textsuperscript{2+} or pH, esp.} \newline
  \mbox{\hsp if site occupancy $<$30\%.} &
  \mbox{$\bullet$ Chips work best $\leq$24 hours after activation.}
  \mbox{$\bullet$ Use higher origami concentration, $\geq 100$~pM.} \newline
  \mbox{\hsp Prepare dilution fresh. Use Lo-Bind tubes.} \newline
  \mbox{$\bullet$ Incubate origami for an hour.} \newline
  \mbox{$\bullet$ If using silanol surface, use $\geq$35~mM Mg\textsuperscript{2+}.} \newline
  \mbox{$\bullet$ If using carboxyl surface, test carboxylation} \newline
  \mbox{\hsp by placing on an unpatterned activated chip.} \newline
  \mbox{$\bullet$ Use pH 8.3--8.5.} 
   \\ \hline 

\vsp  \mbox{High multiple binding.} & 
    Primarily: \newline
    \mbox{\hsp $\bullet$ High origami concentration.} \newline
    \mbox{\hsp $\bullet$ Long incubation time.} \newline
    \mbox{\hsp $\bullet$ Oversized features.} \newline
    \newline
    \newline
    Secondarily: \newline
     \mbox{\hsp $\bullet$ High pH.} \newline
     \mbox{\hsp $\bullet$ High Mg\textsuperscript{2+}.} & 
    First try: \newline
   \mbox{$\hsp \bullet$ Use $\sim$100 pM origami.} \newline
   \mbox{$\hsp \bullet$ Keep incubation between 30 and 90 min.}\newline
   \mbox{$\hsp \bullet$ Look at features in resist by SEM and} \newline
   \mbox{\hsp \hsp adjust e-beam write (feature size, dose)} \newline
   \mbox{\hsp \hsp and/or minimize O\textsubscript{2} activation time.} \newline
    Second try: \newline
     \mbox{$\hsp \bullet$ Keep pH in the range 8.3--8.5.}\newline
     \mbox{$\hsp \bullet$ Use 35~mM Mg\textsuperscript{2+}.} \newline
    \\ \hline 

\vsp \mbox{Poor alignment of origami} \mbox{with few multiple bindings.} & 
     \mbox{$\bullet$ High pH.} \newline
     \mbox{$\bullet$ High Mg\textsuperscript{2+}.} & 
     \mbox{$\bullet$ Keep pH in the range 8.3--8.5.}\newline
     \mbox{$\bullet$ Use $\geq$35~mM Mg\textsuperscript{2+} (if using silanols).} \newline
     \mbox{$\bullet$ Symmetry breaking non-sticky patch is} \newline
     \mbox{ \hsp absent, e.g. poorly written.} \newline  
 \\ \hline 

\vsp \mbox{High background
      binding.} \newline
    \mbox{\hsp $\bullet$ Whole or partial origami} \newline
    \mbox{\hsp \hsp  on background in AFM.} \newline
    \mbox{\hsp $\bullet$ Unstable AFM, {\em e.g.}} \newline
    \mbox{\hsp \hsp  whole scanlines of} \newline
    \mbox{\hsp \hsp identical value (``scars'').} \newline
    \mbox{\hsp $\bullet$ For fluorescent origami,} \newline
    \mbox{\hsp \hsp high background under} \newline
    \mbox{\hsp \hsp optical imaging.}
 & \bnvsp
   \mbox{$\bullet$ Poor initial TMS quality.} \newline
   \mbox{$\bullet$ TMS hydrolyzed by high pH.} \newline
   \mbox{$\bullet$ TMS hydrolyzed by long}\newline
   \mbox{\hsp incubation.} \newline 
   \mbox{$\bullet$ Failure to wash weakly}\newline
   \mbox{\hsp bound origami from TMS.} 
 & \bnvsp
   \mbox{$\bullet$ Dehydrate the wafer by baking before {\em and}} \newline
   \mbox{\hsp after TMS formation.}
   \mbox{$\bullet$ Keep pH$<$9 preferably in the range 8.3--8.5.}\newline
   \mbox{$\bullet$ Keep incubation between 30 and 90 minutes.}\newline
   \mbox{$\bullet$ Remove weakly bound origami with}\newline
   \mbox{\hsp 8$\times$ Tween~20 washes.}\newline
      \\ \hline 

\vsp  \mbox{Large particulates on sites} \mbox{but few or no origami.} & \snvsp
     \mbox{$\bullet$ Sample dewetted or dried.}
   \mbox{\hsp Salts and origami aggregates}
   \mbox{\hsp occupy the site.}  & \snvsp
     \mbox{$\bullet$ Do not let chip dewet during origami}
    \mbox{\hsp deposition or subsequent buffer washes.} \\ \hline 

\vsp  \mbox{Small particles on} \newline
      \mbox{background.}  & 
     \mbox{$\bullet$ Overbaked PMMA.} 
     \mbox{$\bullet$ Acetate causes fine precipitate.}
    & 
     \mbox{$\bullet$ Bake PMMA for 30~s at 180\Tp.} \newline
     \mbox{$\bullet$ Use non-acetate salts/acids when preparing}
     \mbox{\hsp buffers, {\em e.g.} use MgCl\textsubscript{2}, and HCl to adjust.}  \\ \hline 

\vsp  \mbox{Placement requires more} 
    \mbox{than 35 mM Mg\textsuperscript{2+}.} & 
     \mbox{$\bullet$ Surface is too rough}
   \mbox{\hsp or improperly cleaned.}
    & 
     \mbox{$\bullet$ Include HF and NH\textsubscript{4}F cleaning steps.} \newline
     \mbox{\hspace*{1.1in} \small Continues on next page...} \\ \hline

    \end{tabular}
\end{center}

\newpage

\begin{center}
    \begin{tabular}{ | m{4cm} | p{4.5cm} | p{6.5cm} |}
    \hline
    Problem & Likely cause & Solution \\ \hline

\vsp  \mbox{AFM unstable; false engages.} & 
     \mbox{$\bullet$ Tween~20 still present.}
    & 
     \mbox{$\bullet$ Increase buffer washes until surface tension}
     \mbox{\hsp is restored.}  \\ \hline 

\vsp  \mbox{Origami fall off during} 
    \mbox{ethanol drying.} & 
     \mbox{$\bullet$ Too much time spent in}
   \mbox{\hsp dilute ethanol $<$80\%.}
    & 
     \mbox{$\bullet$ Move quickly from low to high \% ethanol.}  \\ \hline 

\vsp  \mbox{Origami ball up into site} \newline
    \mbox{during ethanol drying and} \newline
    \mbox{corners are double height.} \newline
 & \nvsp
     \mbox{$\bullet$ Origami project onto} \newline
   \mbox{\hsp non-sticky TMS surface.} \newline
 & \nvsp
     \mbox{$\bullet$ Hydrolyze TMS surface before drying} \newline
     \mbox{\hsp by incubating in pH~9 buffer.} \newline \\ \hline

    \end{tabular}
\end{center}

\newpage

\label{afm-characterization}
\subsection{AFM characterization}
All AFM images were aquired using a Dimension Icon AFM/Nanoscope V
Scanner (Bruker) using the ``short and fat'' cantilever from an SNL
probe (``sharp nitride lever'', 2~nm tip radius, Bruker).  Non-PCC
samples were imaged in fluid tapping mode, using a cantilever
resonance between 8~and 10~kHz. The use of phase imaging allowed us to
minimize the tip-sample interaction and still achieve high enough
contrast for image analysis. (High-contrast height imaging required
large enough tip-sample forces that origami would occasionally detach
from the surface.) PCC samples were imaged in air in contact mode.
AFM images were processed using Gwyddion ({\tt \small http://gwyddion.net/}).
Single and multiple binding events for placed origami were
hand-annotated and measurements of right triangle and small moon
orientation were made by hand.

\label{toto-optical}
\subsection{TOTO-3 binding and optical experiments}
After placement, small moon origami were labeled with TOTO-3
(Invitrogen/ThermoFisher) and dried via ethanol drying. TOTO-3
labeling was performed by incubating placed origami in a buffer (10~mM
Tris, 35~mM Mg\textsuperscript{2+} at pH~8.3) containing 1~nM TOTO-3
for 10 minutes at room temperature.

All fluorescence imaging was performed with an Olympus BX-61
microscope with a xenon excitation source and Hamamatsu EMCCD cooled
to -75\degree\ C. For fluorescence imaging of simple placed samples
(without PCCs), excitation light was filtered with a 640~nm shortpass
filter and emission light was longpass-filtered via a 645~nm
dichroic. For the PCC array, an additional 655$\pm$5~nm badpass filter
was used to select the PCC's fundamental wavelength of 657.2~nm. For
non-PCC samples, excitation light was filtered with an additional
linear polarizer, mounted on a rotatable adaptor to allow selection of
the desired excitation polarization $\beta$ relative to the sample
axis.  For non-PCC samples, fluorescence emission was collected using
a 50$\times$ objective (1.0~NA oil, optimized for polarized light);
for the PCC array, a 50$\times$ (0.8~NA air) objective was used.

Photoexposure was limited to prevent photobleaching, which could
influence data for which multiple serial images were taken. For both
PCC and non-PCC samples, we observed that complete bleaching took
approximately 45 seconds under constant illumination; we took care to
limit exposure to less than 10\% of this time. For non-PCC samples,
the integration time for each polarization angle was 100
milliseconds. For orientation measurements this meant a total of 3.6
seconds of exposure, for the polarimeter this meant a total of 1.2
seconds of exposure. The final image of the PCC arry (Fig.~4E) was
created by averaging images from three separate samples; each sample
was individually imaged with an integration time of 1~second.

\vsp \Dangersign \shsp \parbox{6in}{Do not label origami with TOTO-3
  prior to placement. Our attempts to label origami with TOTO-3 in
  solution, prior to placement, resulted in no origami binding. This
  is likely due to distortion of the origami's 3D shape upon TOTO-3
  intercalation (which changes DNA twist); profound distortions of DNA
  origami have been observed upon the binding of other intercalators
  ({\em 100,101}). 
By intentionally designing DNA origami with underwinding so that
  intercalated origami have the desired (flat) 3D shape
 ({\em 102}) 
it should be possible to
  achieve placement with origami labelled with TOTO-3 or other
  intercalators.}

\newpage

\begin{figure}[htp]
\includegraphics[width=\textwidth]{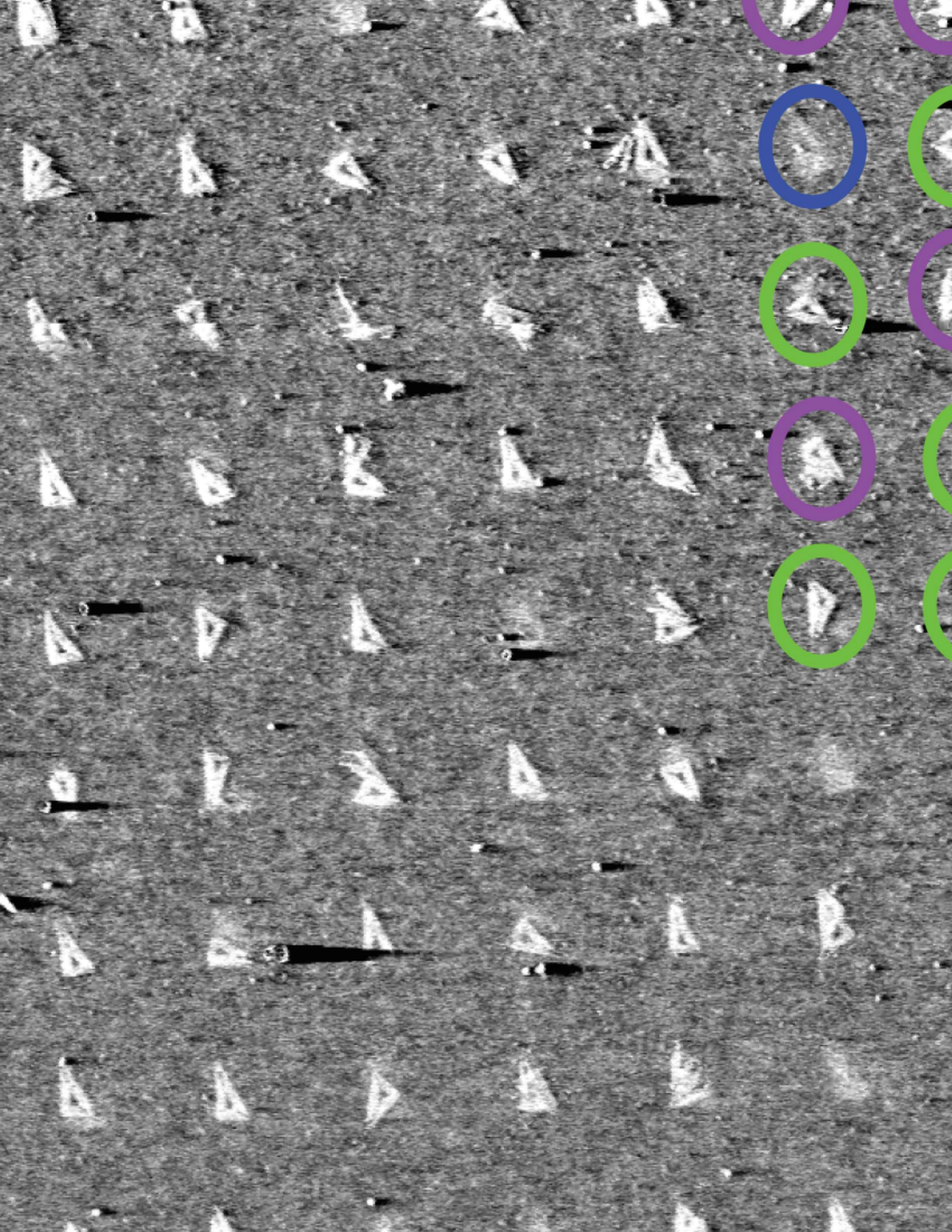}
\caption{{\bf AFM of right triangle origami with its right-hand face
    100\% modified with 20T overhangs, placed on right-handed binding
    sites.} Placement conditions: 100~pM origami, 10~mM Tris, 35~mM
  Mg\textsuperscript{2+}, and pH~8.35 for a 60 minute
  incubation. Ovals give examples of how binding events were
  scored. Red ovals, single origami with roughly the desired
  orientation. Green ovals, single origami with undesired
  orientations. Blue ovals, empty sites. Purple ovals, double bindings
  or other unscored binding events. Angles of single origami relative to
  binding sites (red and green ovals) were measured for 437 sites, to
  the nearest multiple of 4.5\degree.  Scale bar, 2$\mu$m.}
\label{fig:RRTPlaced}
\end{figure}

\begin{figure}[htp]
\begin{center}
\includegraphics[width=4in]{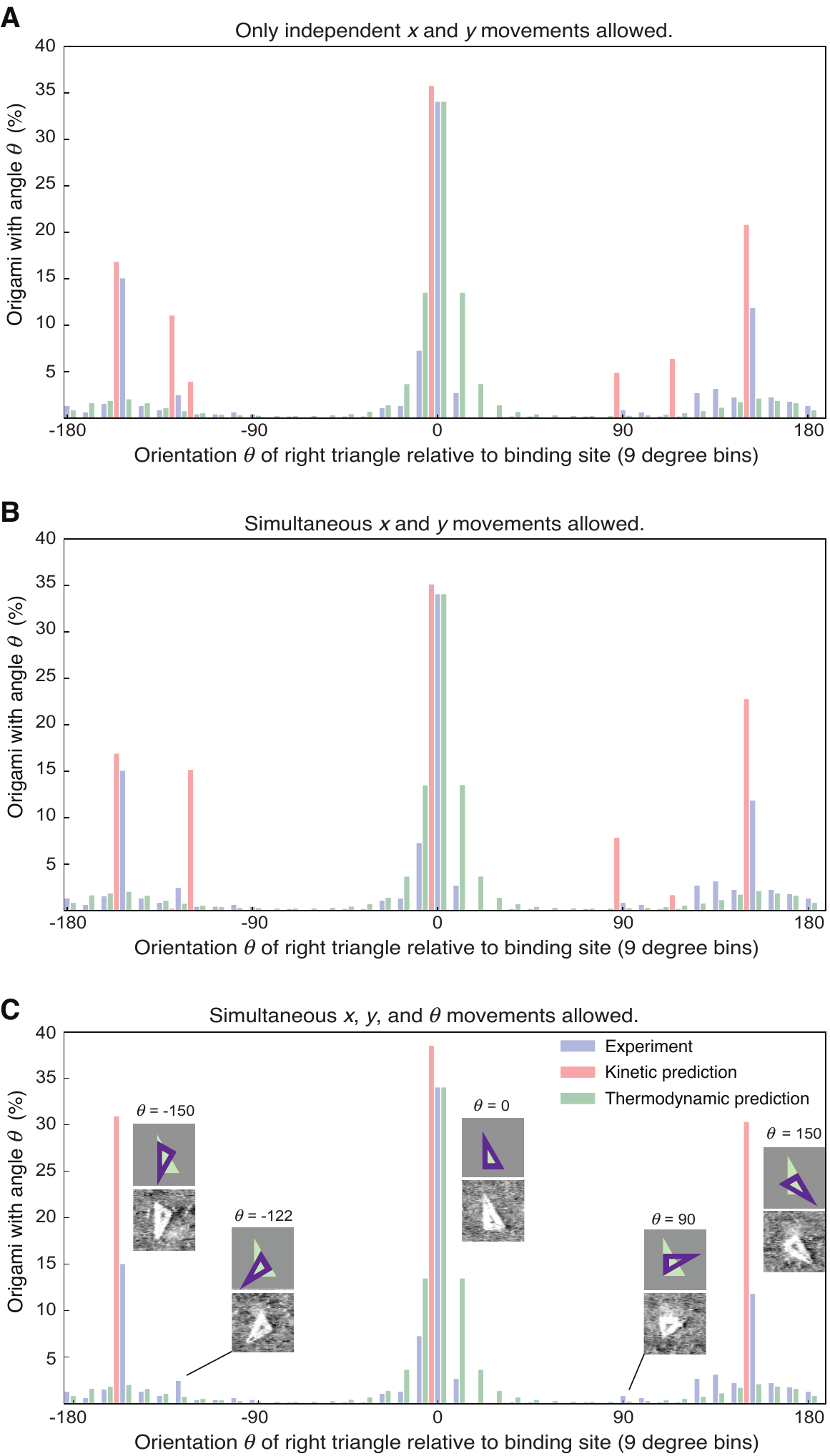}
\end{center}
\caption{{\bf Kinetic simulatins of aligning the right triangle
    origami to its binding site are sensitive to the choice of
    neighborhood in the local move set.}  From any particular
  alignment state ($x$, $y$, $\theta$) its neigborhood is the set of
  states reachable via a single valid transition.  In steepest ascent
  hill climbing, the neighbor that improves the alignment most is
  selected as the new state, and the simulation ends when no neighbor
  can improve upon the current state. ({\bf A}) Neighbors differ from
  current state by $\pm$1~nm in either $x$ or $y$ (but not both), or
  differ by $\pm$1\degree\ in rotation, resulting in 6 neighbors
  total. ({\bf B}) Neighbors differ from the current state by
  translation ($\pm$1~nm in either $x$ or $y$ or both), or any
  rotation ($\pm$1\degree\ rotation), resulting in 10 neighbors total.
  ({\bf C}) Neighbors differ from the current state by any combination
  of translation ($\pm$1~nm in either $x$ or $y$ or both) and rotation
  ($\pm$1\degree\ rotation), resulting in 26 neighbors total.  This
  most restricted neighborhood definition ({\bf A}) results in the
  largest number of local maxima macrostates (7) in the resulting
  state space of the landscape, while the most permissive ({\bf C})
  has the fewest (3). In all cases 1\degree\ microstates were binned
  into 9\degree\ macrostates. Code for these analyses is in {\tt
    \small shapealign-0.1a.tar.gz}}
\label{fig:ThreeModels}
\end{figure}

\begin{figure}[htp]
\begin{center}
\includegraphics[width=5.5in]{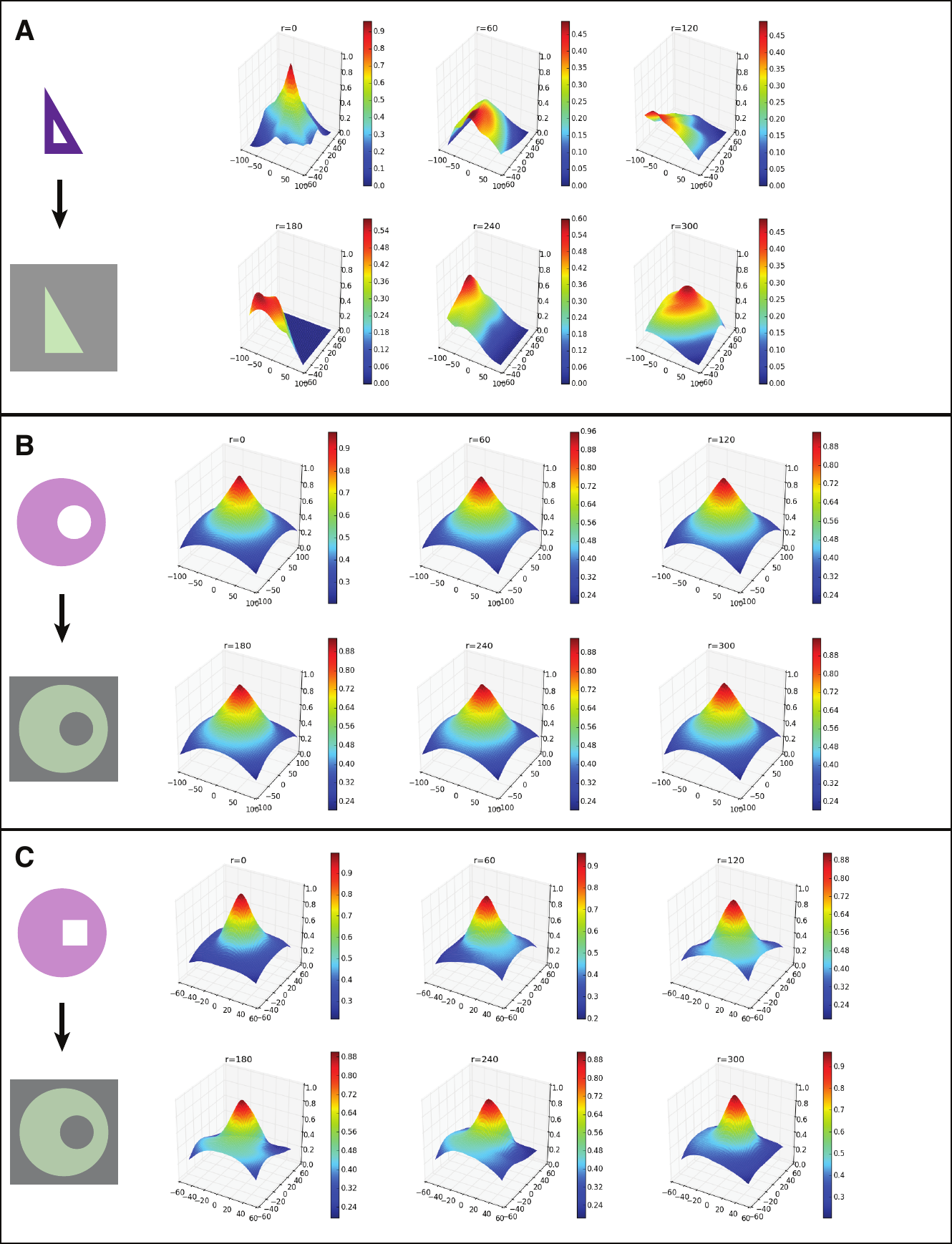}
\end{center}
\caption{{\bf Binding energy landscapes.} Sections of energy
  landscapes for which $x$ and $y$ are allowed to vary, and $\theta$
  is fixed at one of six values, 0\degree, 60\degree, 120\degree,
  180\degree, 240\degree\ and 300\degree.  Colors run from high binding
  energy (red) to low (blue). ({\bf A}) Right triangle binding to a
  right triangle-shaped binding site; the landscape is rugged with
  multiple local maxima. While the right triangle has a hole, the
  binding site does not have a non-sticky region which matches the
  hole. Adding a non-sticky region to match the hole (not shown) does
  not prevent local maxima.  ({\bf B}) Ideal small moon binding to an
  ideal small moon-shaped binding site; the landscape is a smooth
  inverted funnel with a single maxima. The binding site has a
  non-sticky region which matches the hole in the origami; this breaks
  in plane rotational symmetry.  ({\bf C}) Experimental small moon
  binding to an ideal small moon shaped binding site. The shape of the
  hole in the actual experimental origami is a square, rather than a
  circle; the effect is that a single energy maxima is maintained but,
  at certain locations, the slope of the surface is slightly
  flattened.}
\label{fig:LandscapeViews}
\end{figure}

\begin{figure}[htp]
\includegraphics[width=\textwidth]{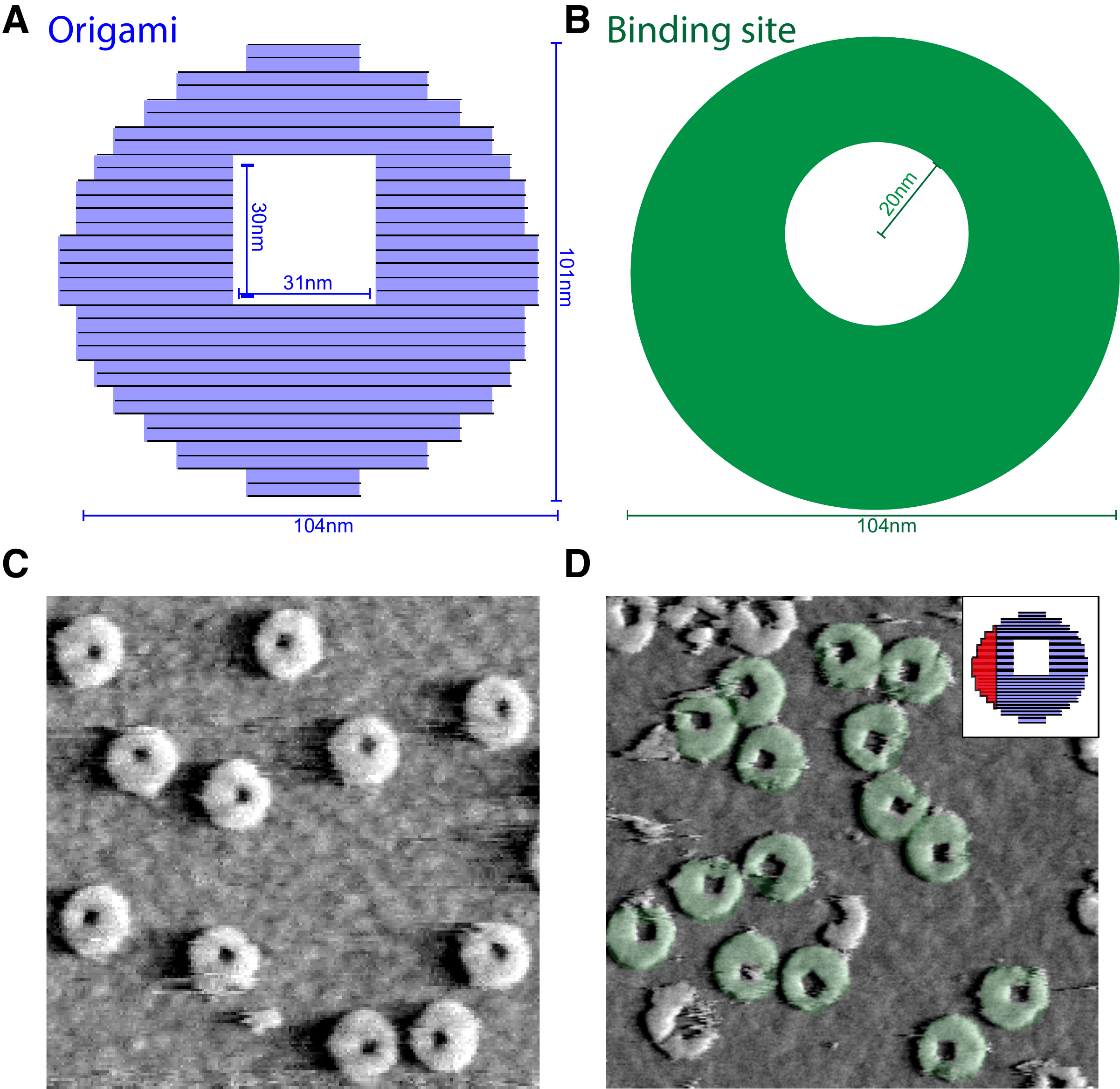}
\caption{{\bf Schematics and AFM for the small moon origami.} ({\bf A)}
  Dimensions of the small moon origami. ({\bf B}) Dimensions of the
  e-beam patterned binding site used for DOP of the small moon
  origami. ({\bf C}) AFM of small moon origami on an unpatterned
  SiO\textsubscript{2} surface. The staples of these small moon
  origami are all modified with 20T extensions, but the symmetry of
  the D1 symmetry of the small moons prevents a determination of
  whether the are landing right-side up (with 20Ts up) or up-side
  down. ({\bf D}) AFM of a modification of the small moon designed to
  help verify that small moons bind SiO\textsubscript{2} right-side
  up. Inset shows a region of staples (red) which were omitted to
  break the D1 symmetry of the small moons. The resulting C1 shape
  allows discrimination based on which edge of the origami looks
  ragged or broken. Green shading indicates origami which were judged
  to be right-side up. Of 642 origami inspected, 95.6\% (614) were
  found to be right-side up; 4.4\% were found to be upside-down or
  their orientation could not be determined.}
\label{fig:DeathStarDesign}
\end{figure}

\newpage

\begin{figure}[htp]
\includegraphics[width=\textwidth]{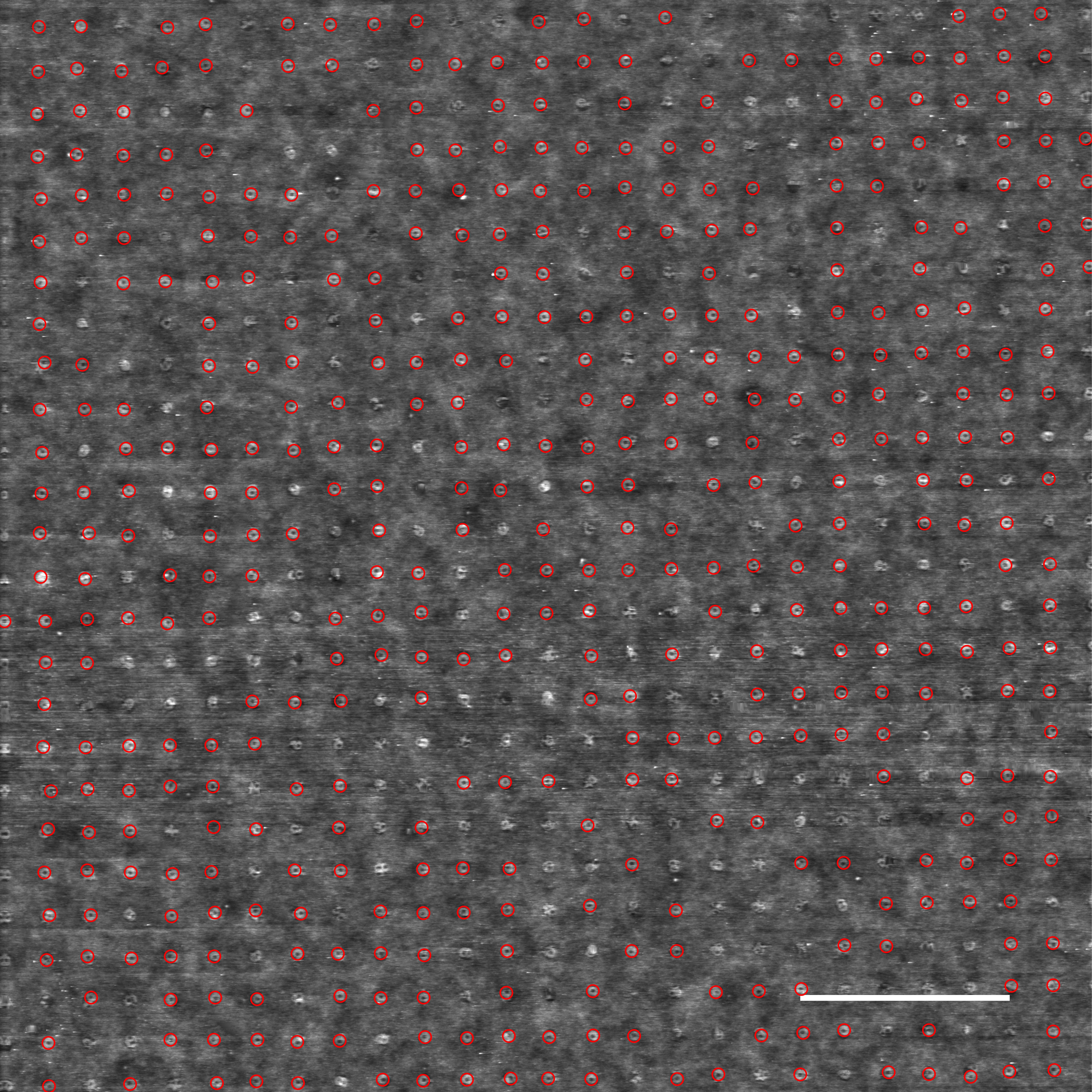}
\caption{{\bf Annotated AFM of small moon origami placed on square
    array of 105~nm diameter disk-shaped binding site.} Scale bar,
  2$\mu$m. Red circles indicate single origami binding events (at 83\%
  of 600 sites) which were cut out automatically and averaged to yield
  annular image in Fig.~2E.}
\label{fig:DeathStarPlacedUnAlignedAnalyzed}
\end{figure}

\begin{figure}[htp]
\includegraphics[width=\textwidth]{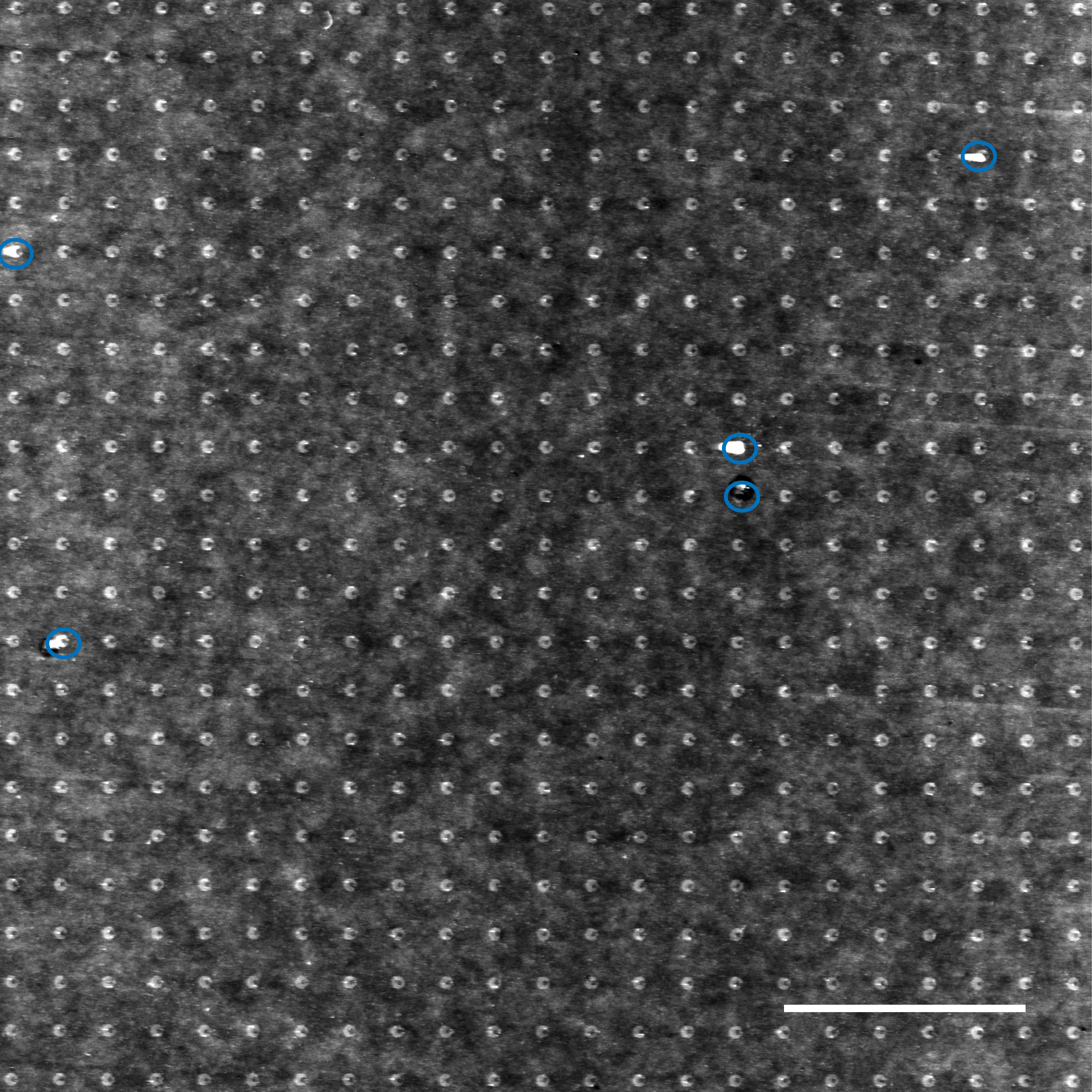}
\caption{{\bf Annotated AFM of small moon origami placed on a square
    array of small moon binding sites.} Scale bar, 2$\mu$m. Blue ovals
  indicate sites which were not analyzed.  The remaining 592 sites
  (98.7\% of 600 total sites; only 529 sites are shown) were cut out
  and averaged to yield the reconstruction of the small moon in
  Fig.~2F. Orientation of each small moon was automatically extracted
  and they were found to be oriented to 0\degree $\pm$6.7 degrees. We
  suggests that the discrepancy between this orientational fidelity,
  and that measured optically ($\pm$3.2\degree) can be explained by a
  poorer ability to measure the orientation of small moons from AFM
  data, which are noisy and have apparent salt artifacts (see white
  dots on origami).}
\label{fig:DeathStarPlacedAlignedAnalyzed}
\end{figure}

\begin{figure}[htp]
\includegraphics[width=\textwidth]{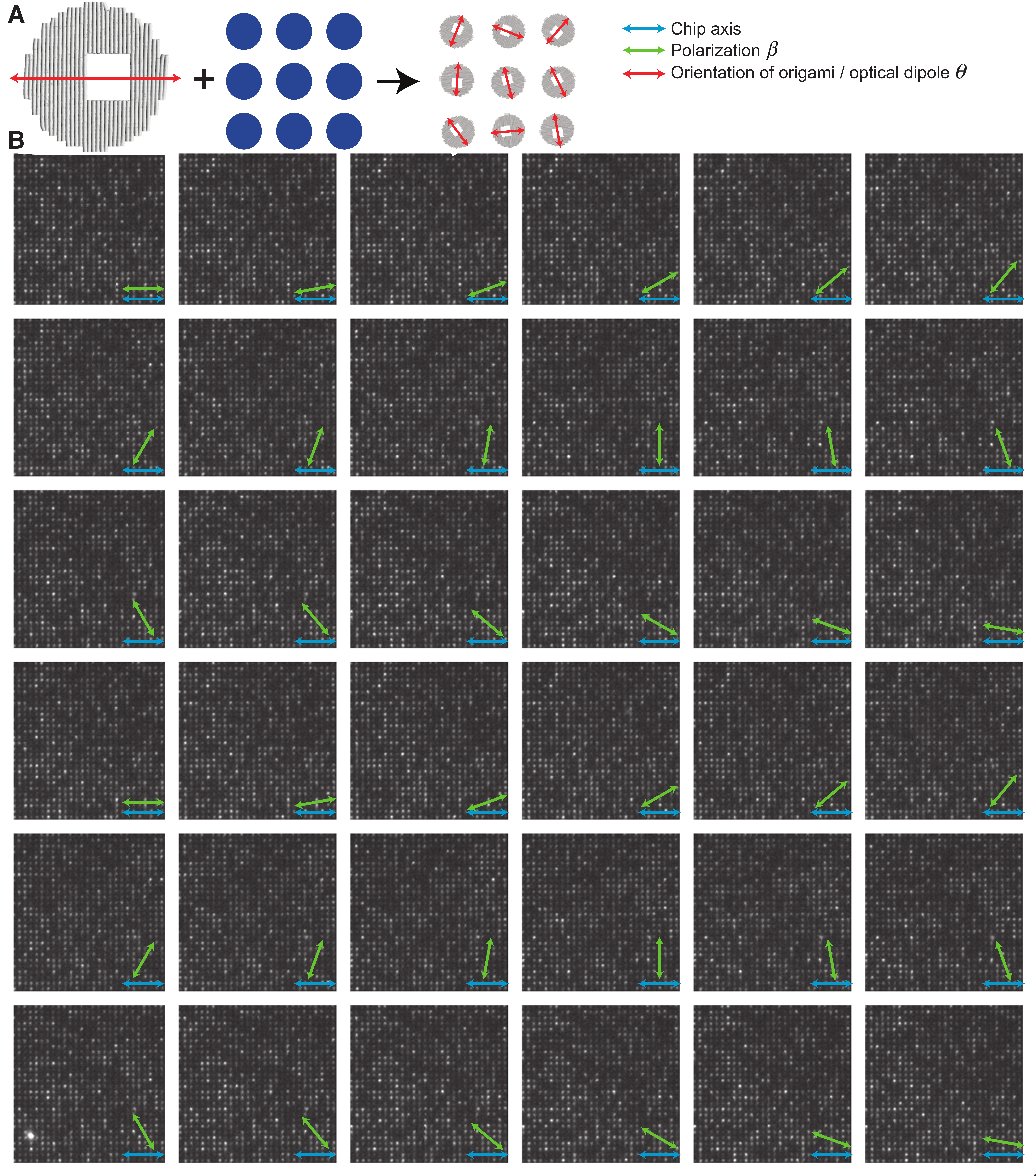}
\caption{{\bf Schematic and raw fluorescence data for small moon
    origami placed on a 1$\bm{\mu}$m period square array 105~nm diameter
    disk-shaped binding sites.} ({\bf A}) Schematic indicates that
  small moons will bind with random orientations and the excitation
  dipoles of intercalated TOTO-3 fluorophores will be uncontrolled.
  ({\bf B}) 36 images show the rotation of excitation light polarization
  (green) relative to the array axis (blue) in 10\degree\ increments.
  Variations in the intensity of small moons is uncorrelated.}
\label{fig:DeathStarPlacedUnAlignedToToMontage}
\end{figure}

\begin{figure}[htp]
\includegraphics[width=\textwidth]{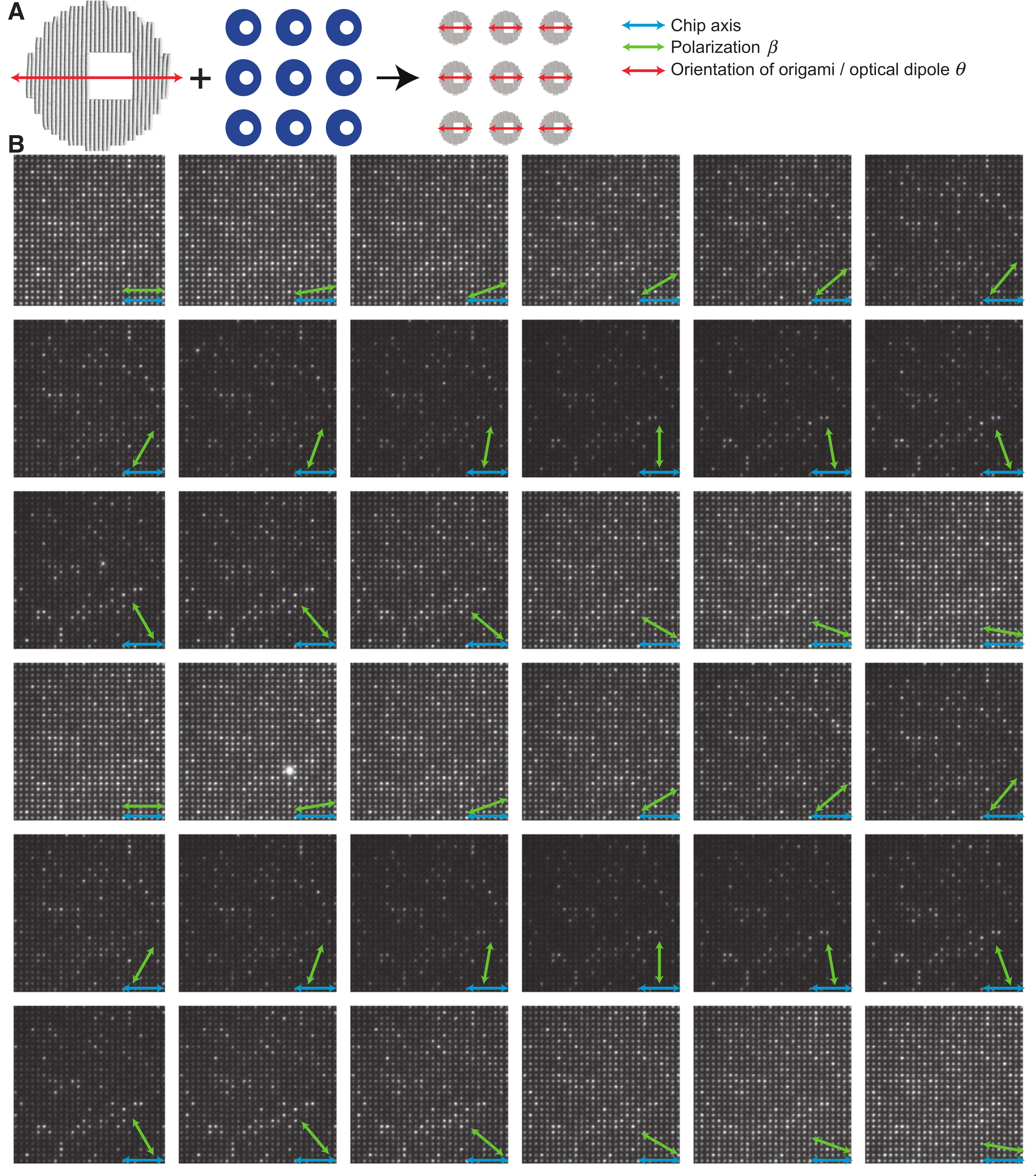}
\caption{{\bf Schematic and raw fluorescence data for small moon
    origami placed on a 1$\bm{\mu}$m period square array of shape-matched
    binding sites.} ({\bf A}) Schematic indicates how the small moon
  origami will align to the binding sites and in turn align the
  excitation dipoles of intercalated TOTO-3 fluorophores. ({\bf B}) 36
  images show the rotation of excitation light polarization (green)
  relative to the array axis (blue) in 10\degree\ 
  increments. Variations in intensity between small moons is highly
  correlated, and small moons are brightest when the polarization axis
  lines up with the array axis.}
\label{fig:DeathStarAlignedToToMontage}
\end{figure}

\begin{figure}[htp]
\includegraphics[width=\textwidth]{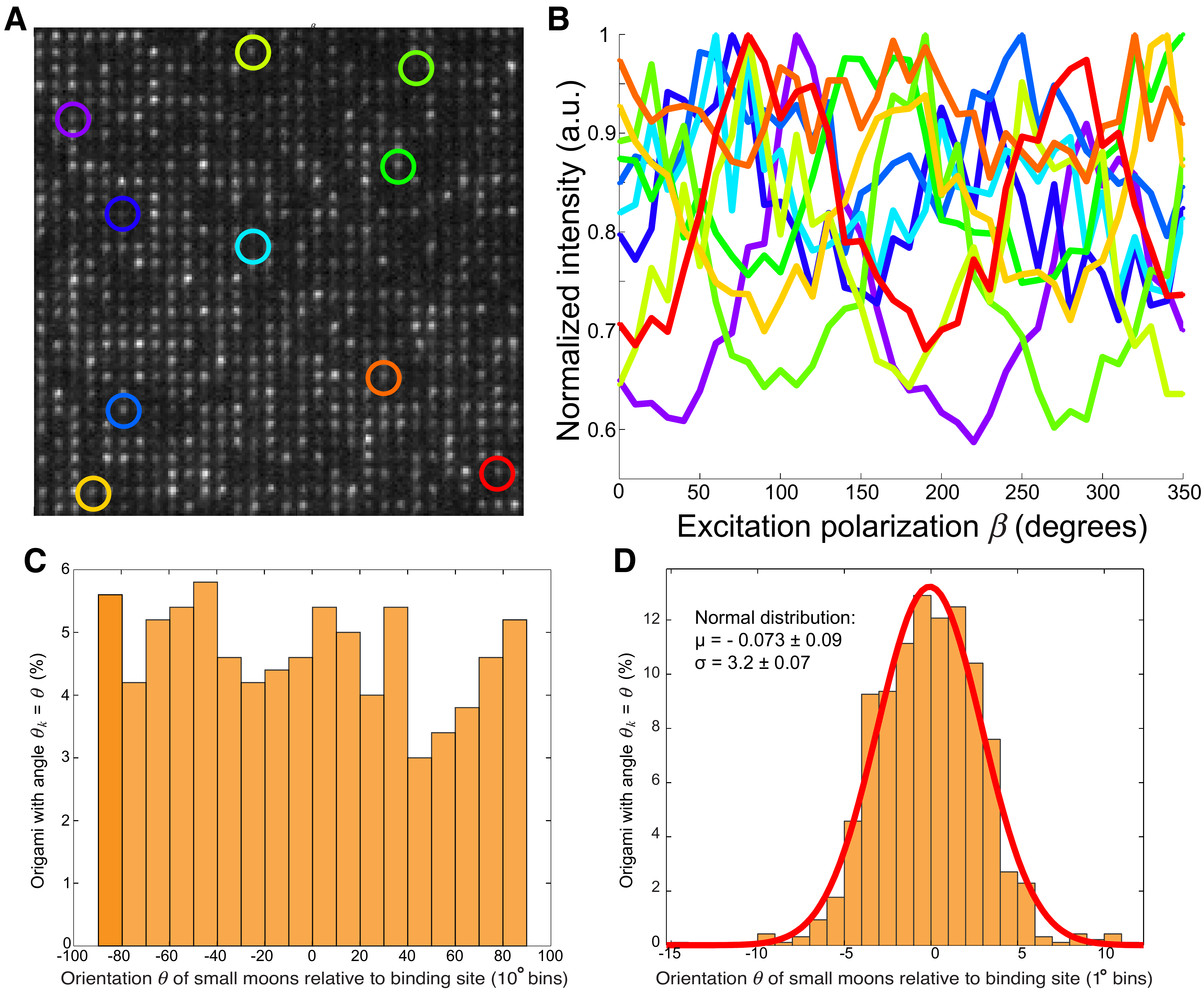}
\caption{{\bf Analysis of orientation based on fluorescence data.}
  ({\bf A}) A subsection of data presented in Fig.~S8, TOTO-3 labelled
  small moons bound to disk-shape binding sites. Ten particular
  binding sites are highlighted with differently colored
  circles. ({\bf B}) Traces of fluorescence intensity from ten binding
  sites highlighted in ({\bf A}), as a function of the orientation of
  excitation polarization $\beta$. All of the $k=1$ to 600 individual
  traces can be fit to $I_{o} \cos^{2}(\beta-\theta_k) + c$. ({\bf C})
  Histogram of $\theta_k$ aggregated into 10\degree\ bins shows that
  the $\theta_k$ are randomly distributed and that small moons exhibit
  no preferential orientation on disk-shaped sites. The flat histogram
  further suggests that the excitation polarization is that intended,
  and that the experimental setup introduces no undesired anisotropy.
  ({\bf D}) Histogram of $\theta_k$ aggregated into 1\degree\ bins for
  data from Fig.~S9, the binding of small moons to shape matched
  binding sites. $\theta_k$ cluster around 0\degree\ with a standard
  deviation of 3.2\degree.}
\label{fig:DeathStarPlacedNotAlignedAnalyzed}
\end{figure}

\begin{figure}[htp]
\includegraphics[width=\textwidth]{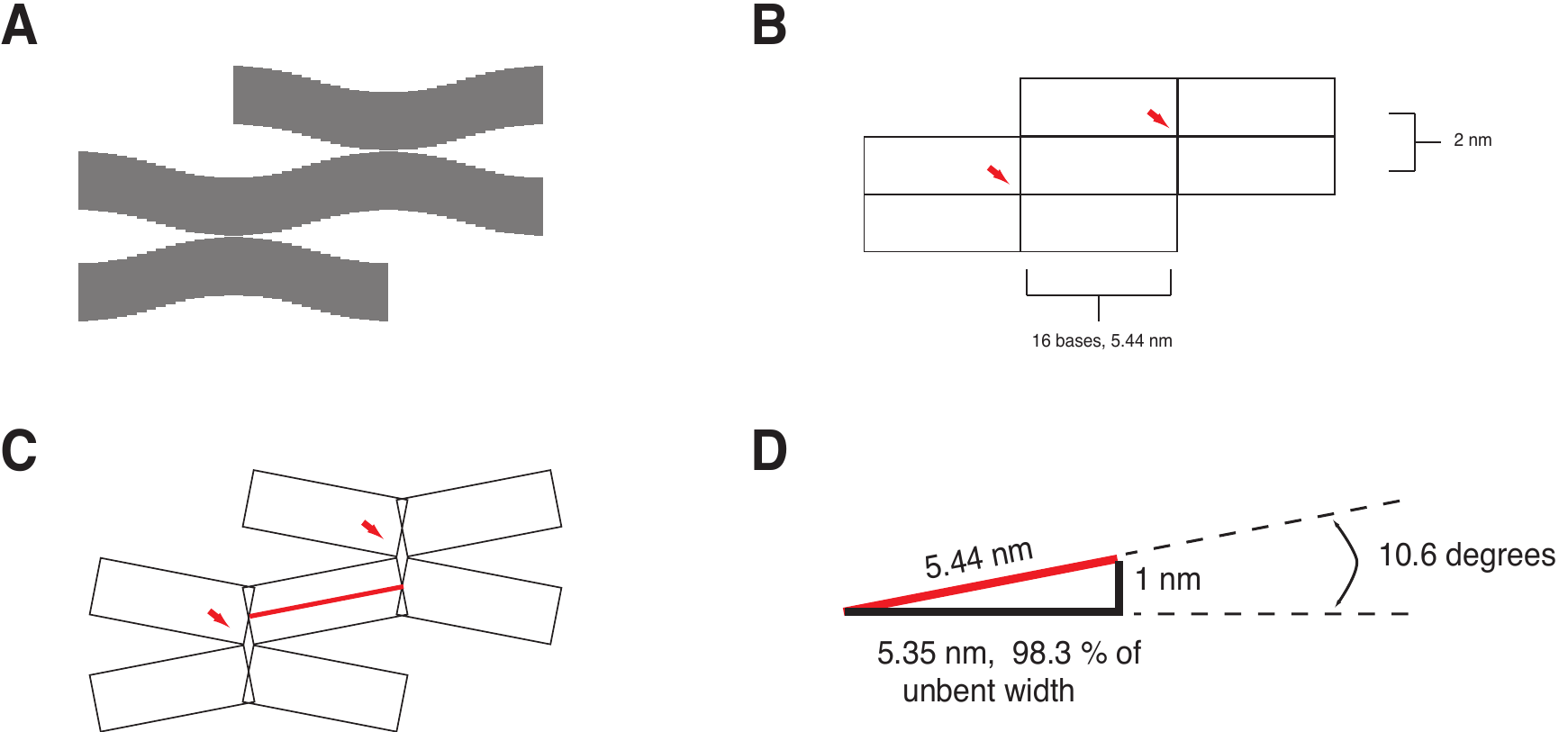}
\caption{{\bf Approximation of helix bending.} ({\bf A}) In DNA
  origami, helices bend back and forth between crossovers, leaving
  gaps. The interhelical gap is empirical. The origami designs used in
  this paper have a 1.5 turn spacing between crossovers, which has
  been consistently observed to add 1~nm of gap per helix. ({\bf B}) A
  model which uses rectangles to depict 1.5 turn, 16 base segments,
  5.44~nm wide and 2~nm tall segments of DNA; here a patch of origami
  with crossovers denoted by red arrows is depicted without
  interhelical bending. ({\bf C}) Coarse grain Monte Carlo
  electrostatic models ({\em 104})
which  capture the interhelical gap result in complex curves which are too
  detailed for estimating helix angle.  Here we simply model helical
  bending using rigid rods, which we allow to overlap slightly at
  cossovers, to yield an average bend angle of 10.6\degree, as
  depicted in ({\bf D}).}
\label{fig:HelixBending}
\end{figure}

\begin{figure}[htp]
\includegraphics[width=\textwidth]{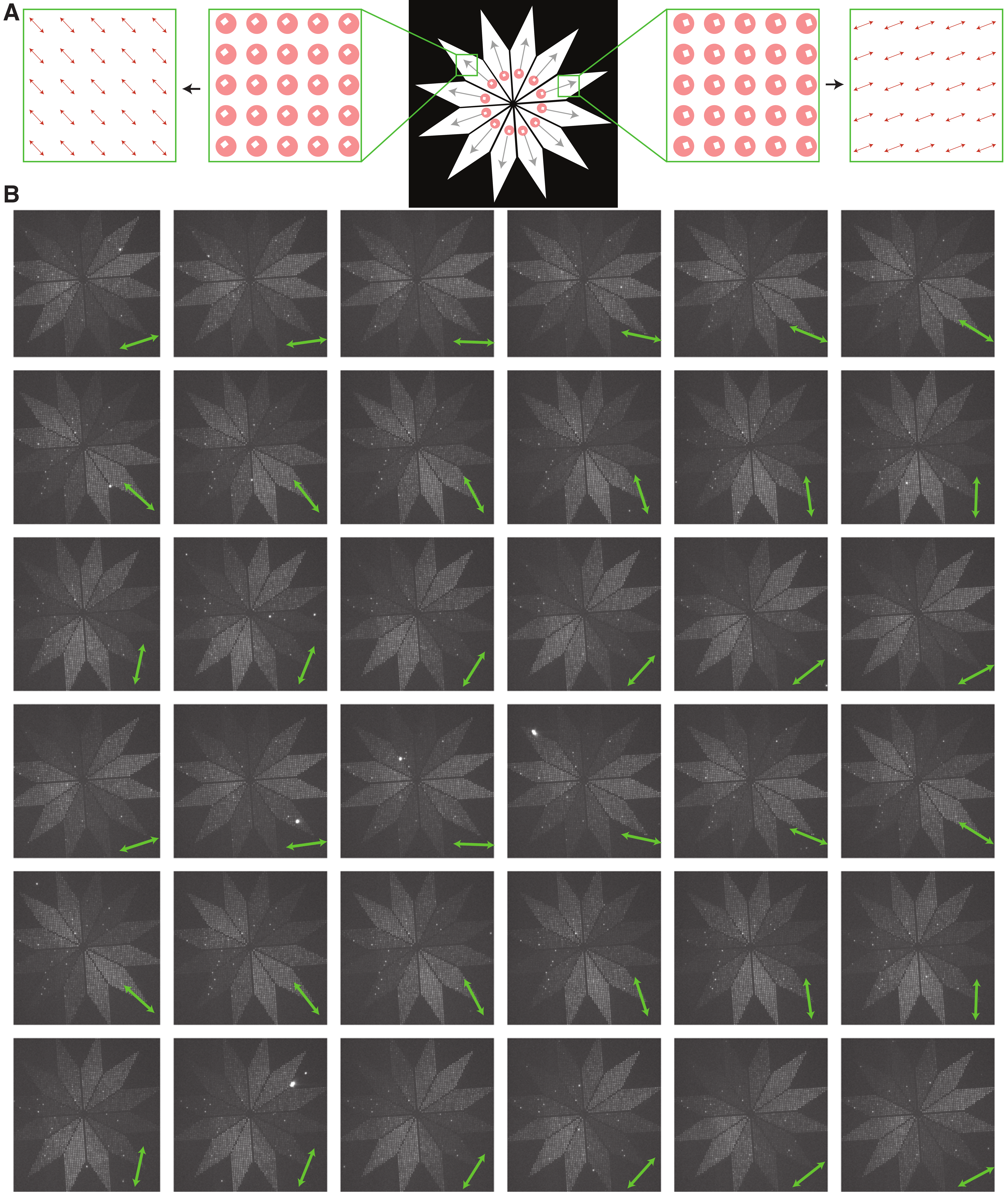}
\caption{{\bf Design and raw data for the polarimeter.} ({\bf A}) Design
  shows the orientation of small moon origami in each of the 12 rays
  of the polarimeter. DNA helices are perpendicular to the ray and so
  the excitation dipole of intercalated TOTO-3 is aligned parallel to
  the ray.  ({\bf B}) 36 images of the polarimeter under polarized
  illumination; green arrows indicate axis of polarization.}
\label{fig:PolarimeterMontage}
\end{figure}

\begin{figure}[htp]
\includegraphics[width=\textwidth]{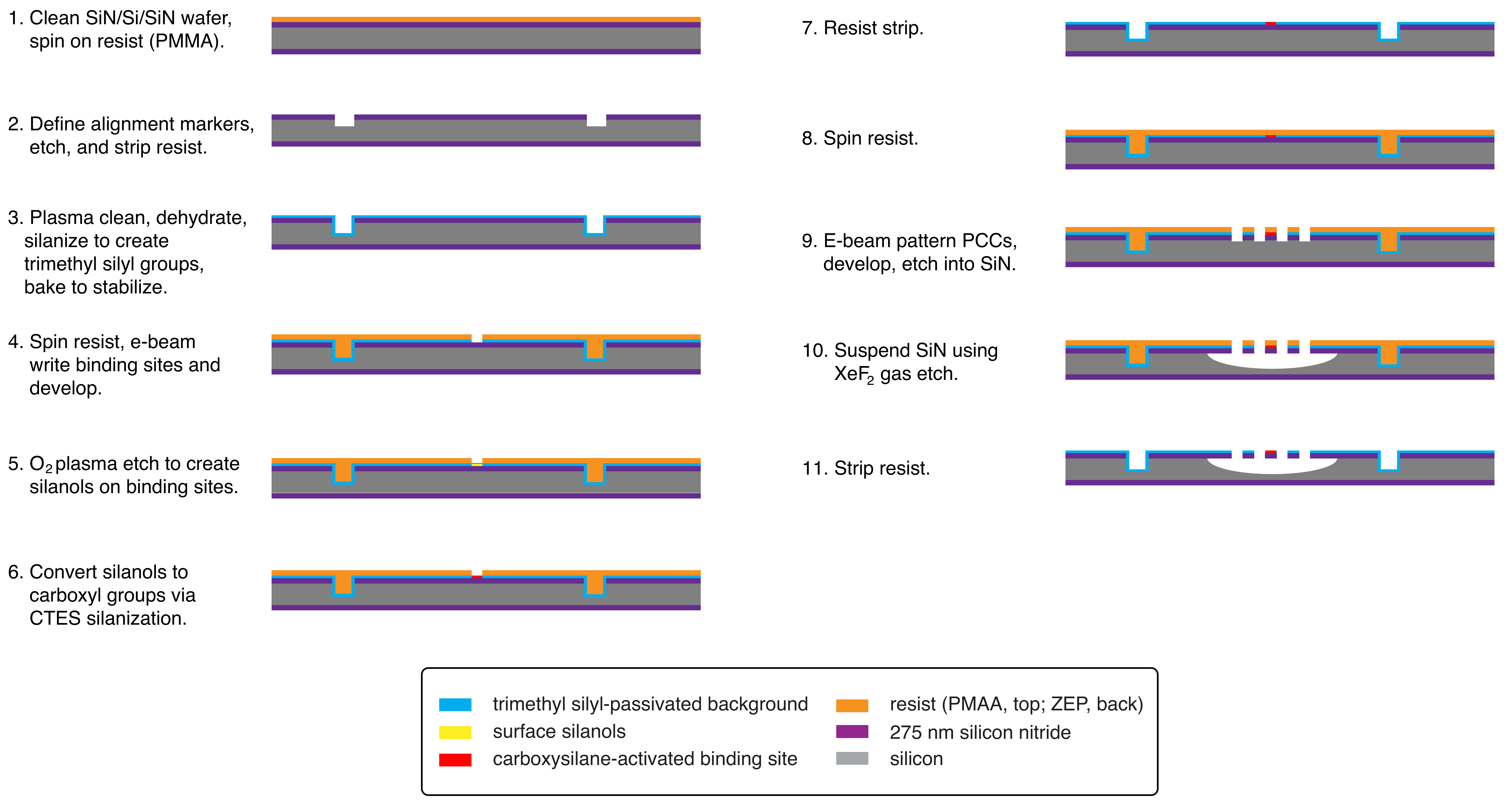}
\caption{ {\bf Process flow for fabricating PCC arrays.}  Note that
  while we used wafers with SiN on both sides, this was just what we
  had available, and wafers with SiN on a single side could have been
  used.  After fabrication, substrates are incubated in origami
  solution, rinsed of excess origami, subject to an ethanol dilution
  series, and air dried.}
\label{fig:PCCfab}
\end{figure}

\begin{figure}[htp]
\includegraphics[width=\textwidth]{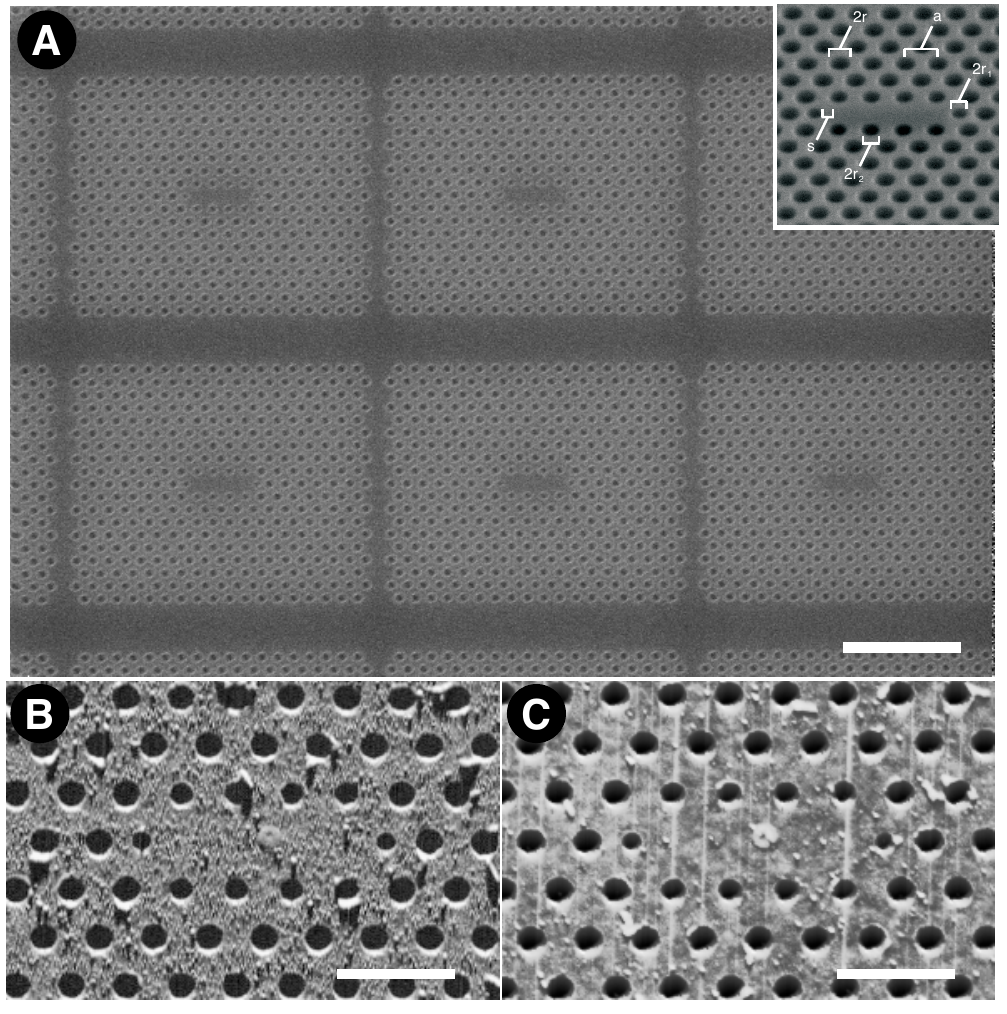}
\caption{ {\bf Photonic crystal arrays for optimizing emitter
    orientation.} ({\bf A}) SEM image of a section of the 13$\times$6
  PCC array; scale bar is 2~$\mu$m. Inset shows critical dimensions of
  different features of the PCC: $a = 256$~nm, $r/a = 0.3$,
  $r_{\mathbf{1}}/a = 0.2$, $r_{\mathbf{2}}/a = 0.25$, $s = 0.22a$.
  ({\bf B}) AFM of a PCC with a single small moon origami oriented
  with its DNA helices parallel to the long axis of the cavity.  ({\bf
    C}) Similar to ({\bf B}), with origami oriented so that its
  helices are perpendicular to the long axis of the cavity.  Scale
  bars for ({\bf B}) and ({\bf C}), 500~nm.}
\label{fig:PCCSEM}
\end{figure}

\clearpage
